\documentclass[aps,pra,pdf,superscriptaddress,twocolumn,showpacs,nofootinbib]{revtex4-2}
\usepackage{amsmath,amssymb,amsfonts}
\usepackage{eqnarray}
\usepackage{graphics,float}
\usepackage{bm}
\usepackage{braket}
\usepackage{amsmath}
\usepackage{physics}
\usepackage{mathtools}
\usepackage{color,xcolor,colortbl}
\usepackage{tikz}
\usetikzlibrary{shapes}
\usepackage[colorlinks=true,
linkcolor=blue,
filecolor=magenta,      
urlcolor=blue,
citecolor=blue]{hyperref}

\definecolor{Gray}{gray}{0.85}
\definecolor{LightCyan}{rgb}{0.25,0.9,0.95}
\definecolor{LightMagenta}{rgb}{0.91,0.21,0.96}

\newcommand{\marker}[2]{\raisebox{0.5pt}{\tikz{\node[draw,scale=0.8,#2,color=#1,fill=#1](){};}}}

\graphicspath{{Figures/}}

\begin{document}
	\title{Two-dimensional miscible-immiscible supersolid and droplet crystal state in a homonuclear dipolar bosonic mixture}
	\author{Soumyadeep Halder}
	\email{soumya.hhs@gmail.com}
	\affiliation{Department of Physics, Indian Institute of Technology Kharagpur, Kharagpur 721302, India}
	
	\author{Subrata Das}
	\email{subratappt@gmail.com }
	\affiliation{Department of Physics, Indian Institute of Technology Kharagpur, Kharagpur 721302, India}
	
    \author{Sonjoy Majumder}
	\email{sonjoym@phy.iitkgp.ac.in}
	\affiliation{Department of Physics, Indian Institute of Technology Kharagpur, Kharagpur 721302, India}
	
	\date{\today}
\begin{abstract}
The recent realization of binary dipolar BEC [\href{https://doi.org/10.1103/PhysRevLett.121.213601}{Phys. Rev. Lett. \textbf{121}, 213601 (2018)}] opens new exciting aspects for studying quantum droplets and supersolids in a binary mixture. Motivated by this experiment, we study groundstate phases and dynamics of a Dy-Dy mixture. Dipolar bosonic mixture exhibits qualitatively novel and rich physics. Relying on the three-dimensional numerical simulations in the extended Gross-Pitaevskii framework, we unravel the groundstate phase diagrams and characterize different groundstate phases. The emergent phases include both miscible and immiscible single droplet (SD), multiple droplets (MD), supersolid (SS), and superfluid (SF) states. More intriguing mixed groundstates may occur for an imbalanced binary mixture, including a combination of SS-SF, SS-MD, and SS-SS phases. We observed the dynamical transition from a miscible MD state to an immiscible MD state with multiple domains formed along the axial direction by tuning the inter-species scattering length. Also by linear quenches of intra-species scattering lengths across the aforementioned phases, we monitor the dynamical formation of supersolid clusters and droplet lattices. Although we have demonstrated the results for a Dy-Dy mixture and for a specific parameter range of intra-species and inter-species scattering lengths, our results are generally valid for other dipolar mixtures and may become an important benchmark for future experimental scenarios.
\end{abstract} 
	
\maketitle

\section{Introduction}
Quantum droplets are dilute liquid-like clusters of atoms produced in a quantum fluid where the dominant attractive mean-field-driven collapse is arrested by the quantum fluctuations \cite{petrov_2015_quantum,lee_1957_eigenvalues}. The supersolid state is also an intriguing state of matter in which the crystalline order of quantum droplets and a global phase coherence \cite{ilzhofer_2021_phase,tanzi_2019_observation} coexist as a result of background superfluid. Both of these states were initially predicted and searched for in liquid helium \cite{toennies_2001_superfluid,toennies_2004_superfluid,barranco_2006_helium,ancilotto_2017_density}. The ability to tune the interaction strength between the particles of an ultracold atomic gas through the Feshbach resonance \cite{chin_2010_Feshbach} offers an excellent platform for studying a plethora of rich physical phenomena. In recent years the quest for quantum droplets and supersolid states in ultracold gases has attracted significant attention. Most theoretical and experimental studies over the last few years reveal the formation of droplets, mainly in two different types of ultracold bosonic systems discussed below.\par 
Quantum droplets in ultracold atomic gases have been first observed in single component dipolar bosonic gases with sufficiently large magnetic dipole moments like dysprosium (Dy) \cite{kadau_2016_observing,wachtler_2016_groundstate,ferrier-barbut_2016_observation}, and erbium (Er) \cite{chomaz_2016_quantumfluctuationdriven,petter_2019_probing,chomaz_2019_longlived}. In this case, when the dipole-dipole interaction (DDI) dominates over the contact interaction in that regime, the anisotropic and long-range characters of DDI lead to the formation of self-bound quantum droplets \cite{schmitt_2016_selfbound,bisset_2016_groundstate,baillie_2017_collective,cinti_2017_classical,baillie_2018_droplet,lee_2021_numerical,wachtler_2016_quantum} and supersolid states \cite{norcia_2021_twodimensional,bland_2022_twodimensional,zhang_2019_supersolidity,roccuzzo_2019_supersolid,poli_2021_maintaining,natale_2019_excitation,hertkorn_2021_pattern,sohmen_2021_birth,leonard_2017_supersolid,tanzi_2019_supersolid,guo_2019_lowenergy,bottcher_2019_transient}. These droplets with highly anisotropic properties have filament-like narrow transverse widths and are elongated along the direction of the external magnetic field.\par 
Quantum droplets also have been realized in non-dipolar binary homonuclear \cite{cabrera_2018_quantum,semeghini_2018_selfbound,cheiney_2018_bright,ferioli_2019_collisions,flynn_2022_quantum} and heteronuclear \cite{guo_2021_leehuangyang,derrico_2019_observation} Bose mixtures. Binary mixtures with an attractive inter-species interaction lead to the formation of miscible droplets. Unlike the droplets formed in a single component dipolar BEC (dBEC) due to the anisotropic and partial attractive nature of DDI, these droplets in a binary system originate solely due to the contact interaction and, therefore, spherical (isotropic) in nature.\par
These phases have been widely explored in various ultracold systems and different experimental setups, ranging from rotating dipolar condensate \cite{zhang_2016_exotic,roccuzzo_2020_rotating,gallemi_2020_quantized,klaus_2022_observation}, dBEC under the influence of a rotating magnetic field \cite{prasad_2019_instability,baillie_2020_rotational,prasad_2021_arbitraryangle,halder_2022_control}, optical lattice trapped dipolar condensate \cite{zhang_2022_supersolid,zhang_2022_supersolid}, lattice trapped atomic mixtures \cite{heidarian_2010_supersolidity,suthar_2020_supersolid,hebert_2008_supersolids}, Rydberg systems \cite{cinti_2010_supersolid,henkel_2012_supersolid}, spin-orbit coupled systems \cite{li_2017_stripe,sachdeva_2020_selfbound}, molecular BECs \cite{schmidt_2022_selfbound}, and a binary mixture of dipolar-nondipolar condensates \cite{li_2022_longlifetime,bland_2022_alternatingdomain}. \par
Recent experimental realization of binary dipolar condensates for the first time \cite{trautmann_2018_dipolar}, and the ability to control their intra-species and inter-species interaction strength through the Feshbach resonance \cite{durastante_2020_feshbach,politi_2022_interspecies} opens new exciting aspects for the study of quantum droplets in a mixture of binary dipolar condensates. Most of the recent theoretical works mainly focus on the formation of a self-bound droplet state in a binary dipolar mixture without any trapping confinement. In contrast to non-dipolar mixtures, due to the anisotropic dipolar counterpart formation of a new class of self-bound miscible, immiscible quantum droplets are predicted \cite{bisset_2021_quantum,smith_2021_approximate,lee_2022_stability,lee_2021_miscibility,smith_2021_quantum,scheiermann_2022_catalyzation}.\par
In this article, we theoretically investigate the possibility of forming different groundstate phases of a binary dBEC (Dy-Dy mixture) confined in a quasi-two-dimensional harmonic trap. For a balanced system, we observe four different groundstate phases: superfluid (SF), supersolid (SS), and single, multiple droplets (SD, MD) that exist in both miscible and immiscible phases. Both components form identical shapes in the miscible regime. Whereas, in the immiscible domain, we observe axially immiscible SD and MD states, and radially immiscible asymmetric SS and SF states. The energetically favored groundstate depends on the number of atoms, intra- and inter-component interactions, and the trap geometry. We depict the phase diagrams and demark all these phases. For an imbalanced mixture, more intriguing states like a mixture of SS-SF, SS-MD, and SS-SS states formed. We have also shown that in an immiscible impurity regime, where one of the components consists of a very small number of atoms (minor component), the major component with a larger number of atoms can bind the impurity component along the axial direction and form a self-bound droplet state for a small intra-species scattering length. Whereas, for comparatively large intra- and inter-species scattering lengths the major component cannot hold the minor component along the axial position. Rather it is pushed along the radially outward direction in presence of the harmonic trapping potential and forms an immiscible mixed state. Using the time-dependent coupled eGPE, we also study the dynamics of a balanced binary system across the above-mentioned phase boundaries.\par
This paper is structured as follows. Section \ref{secii} describes the theory and formalism, including the coupled extended Gross-Pitaevskii equation (eGPE) and the overlap integral to distinguish the miscible and immiscible phases. In Sec. \ref{seciii}, we extract the phase diagrams of the quasi-2D dipolar binary BEC.  Sec. \ref{seciv} characterizes different possible groundstates for an imbalanced binary mixture. In Sec. \ref{secv}, we explore real-time dynamics and the formation of 2D miscible-immiscible droplet and supersolid states by using the time-dependent eGPE. A summary of our findings, together with future aspects, is provided in Sec. \ref{secvi}. Appendix \ref{A} describes the ingredients of our numerical simulations. Appendix \ref{B} is devoted to the variational solution within the same shape approximation (SSA) framework. Appendix \ref{C} delineates the contrast phase diagrams to differentiate the superfluid, supersolid, and droplet phases. In Appendix \ref{D}, we describe the effective potential experienced by one condensate due to the presence of the other condensate. Finally, in Appendix \ref{E}, we have shown the time evolution of density profiles and the overlap integral following an interaction quench of a miscible SF state across the relevant phase boundaries.

\section{Theory}\label{secii}
We consider a mixture of two species of dipolar bosonic atoms with a large magnetic dipole moment $\mu_i^m$ $(i=1,2)$ polarized along the $z$ direction by an external magnetic field and confined in a circular symmetric harmonic trapping potential. In the ultracold regime, the atoms of species-$i$ are characterized by the macroscopic wave function $\psi_i$, whose temporal evolution is described by the coupled eGPE:
\begin{align}
	i\hbar\frac{\partial\psi_i(\textbf{r},t)}{\partial t}&=\Big[-\frac{\hbar^2}{2m_i}\nabla^2+V_t(\textbf{r})+\sum_{j=1}^{2}\Big(g_{ij}\abs{\psi_j(\bf{r},t)}^2+\nonumber\\&\int d\textbf{r}' V_{ij}^{dd}(\textbf{r}-\textbf{r}')\abs{\psi_j(\bf{r}^{\prime},t)}^2\Big)+\Delta \mu_i\Big]\psi_i(\textbf{r},t).\label{egpe}
\end{align}
Here, $V_t(\textbf{r})=\frac{1}{2}m_i\omega^2(x^2+y^2+\lambda^2z^2)$ is the harmonic trapping potential with angular frequencies $\omega_x=\omega_y=\omega,\omega_z$; $m_i$ is the atomic mass of the $i$'th species and $\lambda=\omega_z/\omega$ is the trap aspect ratio. The short-range intra- and inter-component interaction strengths are given by $g_{ii}=4\pi\hbar^2a_{ii}/m_i$ and $g_{ij}=2\pi\hbar^2a_{ij}/m_{ij}$, respectively. Here, $a_{ii}$ and $a_{ij}$ are the intra- and inter-component scattering length of atoms and $m_{ij}=m_im_j/(m_i+m_j)$ is the reduced mass. Apart from the contact interaction, there exists a long-range DDI between the atoms, and it takes the form
\begin{equation}
	V_{ij}^{dd}(\textbf{r})=\frac{3g_{ij}^{dd}}{4\pi}\left(\frac{1-3\cos^2\theta}{r^3}\right),\label{dip_pot}
\end{equation}
where $g_{ij}^{dd}=2\pi\hbar^2a_{ij}^{dd}/m_{ij}$ is the DDI strength between the atoms of $i$'th and $j$'th species, with the DDI length $a_{ij}^{dd}=\mu_0\mu_i^m\mu_j^m m_{ij}/6\pi\hbar^2$, and $\theta$ is the angle between the axis linking the two particles and the dipole polarization direction ($z$-axis). The last term appearing in Eq.({\ref{egpe}}) represents the correction to the chemical potential resulting from the effect of quantum fluctuation given by \cite{bisset_2021_quantum,smith_2021_approximate,smith_2021_quantum}
\begin{equation}
	\Delta \mu_i=\frac{m_i^{3/2}}{3\sqrt{2}\pi^2\hbar^3}\sum_{\pm}\int_{0}^{1}\dd u~\Re{I_{i\pm}},
\end{equation}
where
\begin{align}
	I_{1\pm}=\bigg(\tilde{U}_{11}\pm &\frac{\delta_1 \tilde{U}_{11}+2\tilde{U}_{12}^2 n_2}{\sqrt{\delta_1^2+4\tilde{U}_{12}^2n_1 n_2}}\bigg)\bigg(n_1\tilde{U}_{11}+\nonumber\\& n_2\tilde{U}_{22}\pm\sqrt{\delta_1^2+4\tilde{U}_{12}^2n_1 n_2}\bigg)^{3/2},
\end{align}
with $\delta_1=n_{1}\tilde{U}_{11}-n_2\tilde{U}_{22}$, and $\tilde{U}_{ij}(u)=g_{ij}[1+\epsilon_{ij}^{dd}(3u^2-1)]$, being the Fourier transform of the total interaction potential and the dimensionless parameter $\epsilon_{ij}^{dd}=a_{ij}^{dd}/a_{ij}$, quantifies the relative strength of DDI to the contact interaction between the atoms in species-$i$ and $j$. A similar expression for $\Delta\mu_2$ can be easily obtained with $\delta_2=-\delta_1$.  The order parameters of each of the condensates are normalized to the total number of atoms in that species $N_i=\int \dd\textbf{r}\abs{\psi_i(\textbf{r})}^2$.

\subsection{Overlap integral}
A binary dBEC can exhibit a miscible or immiscible phase. A well-known measure to characterize these two phases is the overlap integral, defined as
\begin{equation}
	\Lambda=\frac{\left[\int d\textbf{r}n_1(\textbf{r})n_2(\textbf{r})\right]^2}{\left[\int d\textbf{r}n_1^2(\textbf{r})\right]\left[\int d\textbf{r}n_2^2(\textbf{r})\right]},\label{lambdao}
\end{equation}
where $n_i(\textbf{r})=\abs{\psi_i(\textbf{r})}^2$ is the densities of the species-$i$. $\Lambda=1$ implies maximal spatial overlap between the condensates, i.e., the system is in a completely miscible state. Whereas, a complete phase separation (immiscible phase) corresponds to $\Lambda=0$.


\section{Groundstate phases of a balanced mixture}\label{seciii}
To illustrate the groundstate properties and explore different phases of a Dy-Dy mixture\footnote{We consider both the species have equal mass m. This is a good approximation for the mixture of $^{162}$Dy, $^{164}$Dy (suitable for experiments) with a relative difference between mass extremes of less than 2\%.}, we find that the intriguing parameters are the intra- and inter-component scattering lengths $(a_{ii}, a_{ij})$, the trap aspect ratio ($\lambda$) and the number of atoms in the condensate ($N_i$). Here, we consider a balanced mixture with equal intra-species interactions ($a_{11}=a_{22},~a_{11}^{dd}=a_{22}^{dd}$) and an equal number of particles in each species ($N_1=N_2=N$). We first evaluate the groundstate of the binary mixture as a function of intra- and inter-component scattering lengths and the number of atoms in the species $i=1,2$ keeping the trap aspect ratio fixed at $\lambda=2.95$. Subsequently, we also investigate the effect of trap geometry on the groundstate phases by varying the trap aspect ratio with the intra-species scattering length for a fixed number of particles and inter-species scattering length.\par
 A binary mixture can be either in a miscible or immiscible phase. We differentiate the miscible and immiscible phases by numerically evaluating the overlap integral $\Lambda$ (Eq. (\ref{lambdao})). In the large $N$ limit, the effect of quantum pressure is negligible compared to the non-linear interactions, and the condensate can be well approximated by the Thomas-Fermi (TF) approximation. Thus immiscibility is completely determined by the intra- and inter-component scattering lengths for a balanced system (where we can apply SSA, see the Appendix \ref{B}) and the transition occurs when $a_{12}=\sqrt{a_{11}a_{22}}$. However, when both condensates consist of a small number of particles, quantum pressure makes a significant contribution to the condition of immiscibility transition. Quantum pressure of individual species is $P_i=\frac{\hbar^2}{2m_i \sqrt{n_i}}\nabla^2\sqrt{n_i}$, where $n_i$ is the density of the species-$i$. This pressure describes the attractive force due to spatial variation of density, which becomes maximum at the interface when the two condensates are in an immiscible phase. As a consequence, to minimize the quantum pressure energy for a small number of particles, the miscible to immiscible transition boundary deviated from $a_{12}=\sqrt{a_{11}a_{22}}$ and the binary system favors the miscible state more, as can be seen in Fig. \ref{fig:1}(a) and \ref{fig:1}(b).\par 
Due to the anisotropic DDI, the SF, SS, and droplet (SD and MD) phases emerge in a dBEC. These phases are best characterized by the density contrast $\mathcal{C}=(n_{\rm max}-n_{\rm min})/(n_{\rm max}+n_{\rm min})$ \cite{bland_2022_alternatingdomain}, where $n_{\rm max}$ and $n_{\rm min}$ are the neighbouring maximum and minimum densities as one moves on the $x-y$ plane (a plane perpendicular to the polarization direction). This allows us to depict different phase domains in the phase diagrams, where we take $\mathcal{C}=0$ to be a superfluid phase, and consider $0<\mathcal{C}<0.96$ to be a supersolid and $\mathcal{C}>0.96$ as a droplet state \cite{bland_2022_alternatingdomain}. For a detailed discussion on the density contrast see Appendix \ref{C}.\par


\subsection{Phsae diagrams of binary dipolar condensate}
\subsubsection{Intra-species scattering length $(a_{ii})$ vs. population $(N)$}
\begin{figure}[tb!]
	\centering
	\includegraphics[width=0.48\textwidth]{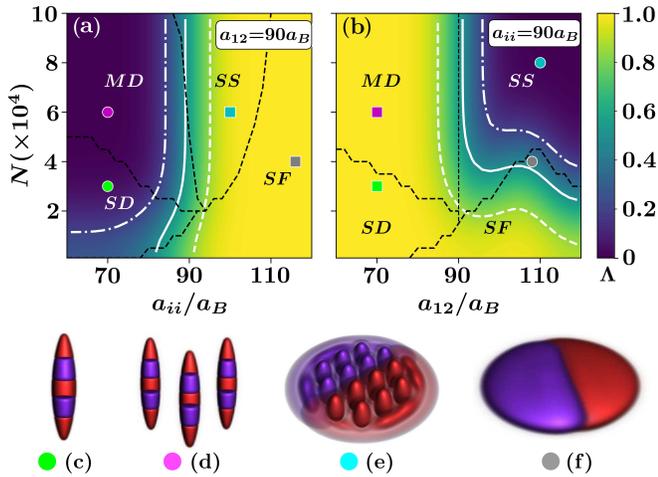}
	\caption{Groundstate phase diagrams in an oblate harmonic trap of a balanced Dy-Dy mixture as a function of the number of particles and (a) intra- and (b) inter-component s-wave scattering length. The colorbar corresponds to the value of overlap integral $\Lambda$. The white solid contour has been drawn at $\Lambda=0.5$ and it indicates the phase transition from a miscible to an immiscible groundstate. The other two white dashed and dash-dotted lines are drawn at $\Lambda=0.9$ and $\Lambda=0.1$, respectively. Whereas the black dashed lines separate the phase diagram in superfluid (SF), supersolid (SS), and single, multiple insulating droplets (SD, MD) states. Some examples of immiscible 3D isosurfaces of (c) SD (\protect\marker{green}{circle}), (d) MD (\protect\marker{LightMagenta}{circle}), (e) SS (\protect\marker{LightCyan}{circle}), and (f) SF (\protect\marker{gray}{circle}) states are highlighted by the symbols demarked in the groundstate phase diagrams (a) and (b).  Results are for the case of $N_1=N_2=N$, (a) $a_{12}=90a_B$, (b) $a_{ii}=90a_B$ with $a_{ii}^{dd}=131a_B ~(\mu_i^m=9.93\mu_B)$ where $a_B$ is the Bohr radius, and $(\omega_x,\omega_y,\omega_z)=2\pi\times(45,45,133)$Hz.}\label{fig:1}
\end{figure}

\begin{figure}[tb!]
	\centering
	\includegraphics[width=0.48\textwidth]{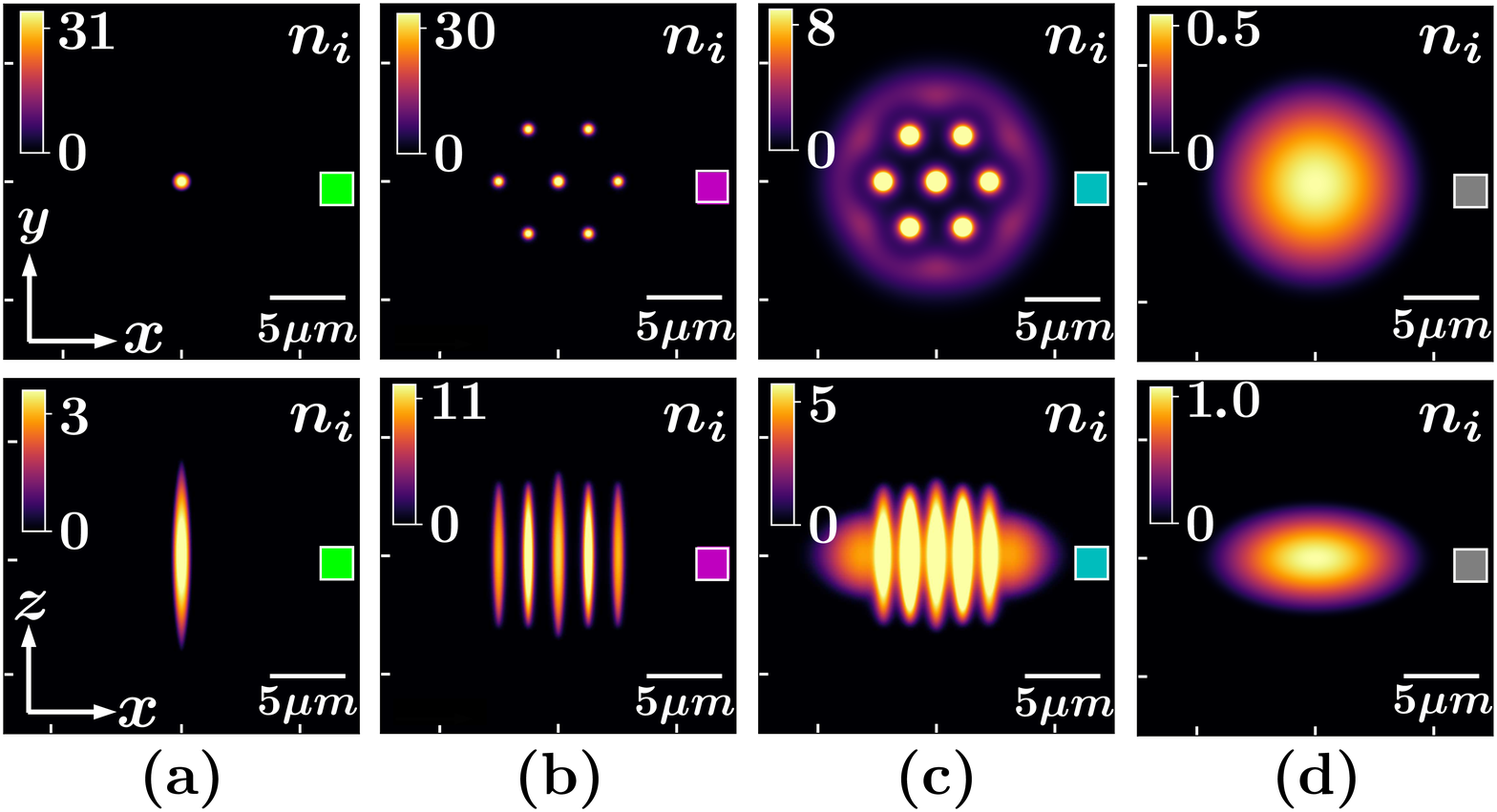}
	\caption{Groundstate density profiles of a miscible (a) SD (\protect\marker{green}{rectangle}), (b) MD (\protect\marker{LightMagenta}{rectangle}), and (c) SS (\protect\marker{LightCyan}{rectangle}), and (d) SF (\protect\marker{gray}{rectangle}) states, are highlighted by the corresponding marker demarked in the groundstate phase diagrams (see Fig.\ref{fig:1}(a), (b)). $n_i$ is the density of the species-$i$, where $i= 1,2$ (both the components have identical shapes). The first and second row represents the integrated density profile in the $xy$ ($n_i(x,y)=\int \dd z~ n_i (x,y,z) $) and $xz$ ($n_i(x,z)=\int \dd y~ n_i (x,y,z) $) planes, respectively. Colorbar represents the density of each species in units of $10^3\mu m^{-2}$. Results are for the case of a Dy-Dy mixture with $N_1=N_2=N,~a_{11}=a_{22},~a_{ii}^{dd}=131a_B~(\mu_i^m=9.93\mu_B),$ and $\{\omega_{x},\omega_y,\omega_{z}\}=2\pi\times\{45,45,133\}$Hz. Other parameters are (a) $a_{ii}=90a_B,~a_{12}=70a_B,~N=3\times10^4$, (b) $a_{ii}=90a_B,~a_{12}=70a_B,~N=6\times10^4$, (c) $a_{ii}=100a_B,~a_{12}=90a_B,~N=6\times10^4$ and (d) $a_{ii}=116a_B,~a_{12}=90a_B,~N=4\times10^4$.}\label{fig:Miscible_SD_MD_SS_2d_den}
\end{figure}

 Here we construct a groundstate phase diagram with the intra-species scattering length $a_{ii}~(i=1,2)$ and the number of particles $N_1=N_2=N$ for a constant inter-species scattering length $a_{12}$ (see Fig. \ref{fig:1}(a)). To demonstrate the phase diagram, we fix the inter-species scattering length at $a_{12}=90a_B$ and vary the intra-component scattering length $a_{ii}$ from $60a_B$ to $120a_B$, and the number of atoms $N$ from $10^3$ to $10^5$ of each species. The balanced binary mixture remains in a miscible phase for a large value of $a_{ii}(>a_{12})$. Miscible to immiscible transition for large number of particles, $N > 4\times10^4$, occurs at $a_{ii}=90 a_B$ $(a_{12}=\sqrt{a_{11}a_{22}})$. However as mentioned above, for $N< 4\times 10^4$, this transition occurs at smaller $a_{ii}(<90a_B)$. This transition is indicated by the solid white line corresponding to $\Lambda=0.5$ in Fig. \ref{fig:1}(a). For sufficiently large $a_{ii}$, due to the dominated short-range contact type interactions over the DDI, the binary dipolar mixture remains in a miscible SF state. It corresponds to a smooth (non-modulated) quasi 2D TF density distribution with a low peak density (see Fig. \ref{fig:Miscible_SD_MD_SS_2d_den}($\rm d$)). As we decrease $a_{ii}$ down to a critical value, each component of the mixture undergoes an abrupt phase transition to a 2D SS state (overlapping droplets) for a sufficiently large number of particles $(N>2\times 10^4$). These droplets are coupled via a low-density superfluid. In this regime, we get two coexisting miscible SS states (Fig. \ref{fig:Miscible_SD_MD_SS_2d_den}(c)) due to $a_{ii}>a_{12}$. In contrast, for a small number of particles $N<2\times 10^4$, no droplet nucleation is observed in this regime, and both the components of the binary mixture remain in a miscible superfluid state. When further decreasing $a_{ii}$ below $90a_B$, two components become immiscible due to comparatively large inter-component scattering length $(a_{12}>a_{ii})$ and the density overlap between the droplets vanishes rapidly. In this sufficiently low $a_{ii}$ regime to minimize the DDI energy, atoms of each species form multiple separate domains along the axial direction and the binary mixture forms a multi-domain droplet state. Since we have taken a balanced mixture with equal intra- and inter-species interaction strength, the binary mixture forms a symmetric immiscible droplet state. For a small number of particles, we observed an immiscible single droplet state (SD) (Fig. \ref{fig:1}(c)). In the case of a sufficiently large number of particles, we obtain an immiscible multiple droplet state (MD) (Fig. \ref{fig:1}(d)).\par

\subsubsection{Inter-species scattering length $(a_{12})$ vs. population $(N)$}
Now, in case of a fixed intra-component scattering length $(a_{ii}=90a_B)$, we construct a groundstate phase diagram (see Fig. \ref{fig:1}(b)) by varying the inter-species scattering length $a_{12}$ and the number of atoms $N$ in each species. For a sufficiently large $a_{12} (>80a_B)$ and a small number of particles, the stationary state solution of the dipolar mixture is a miscible SF state. The increase in the number of particles $N$ induces a transition to an immiscible SF regime. In this case, since we have taken a balanced mixture, there is no preference over which one particular component remains at the center. So the groundstate of the balanced binary mixture has one domain of each species and is separated along the $x-y$ plane (radial direction), producing an asymmetric immiscible SF state (see Fig. \ref{fig:1}(f)). As we further increase the number of particles ($N>4\times 10^4$), the smooth non-modulated density profile of each domain undergoes a phase transition and each species develops a periodic density modulated pattern along the $x-y$ plane. The density humps (droplets) are connected by lower-density regions (superfluid). Both species unveil SS properties. However as we discussed above, in this phase regime due to large $a_{12}~(>a_{ii})$, the phase of the binary mixture is radially separated and we obtain an asymmetric immiscible SS state (Fig. \ref{fig:1}(e)). At a lower $a_{12}(<a_{ii})$, the density overlap between the droplets in each species vanishes completely. Furthermore, depending on the number of particles, the binary system displays a miscible SD (small number of particles) and MD (large number of particles) state as portrayed in Fig. \ref{fig:Miscible_SD_MD_SS_2d_den}(a) and \ref{fig:Miscible_SD_MD_SS_2d_den}(b). 

\begin{figure}[tb!]
	\centering
	\includegraphics[width=0.48\textwidth]{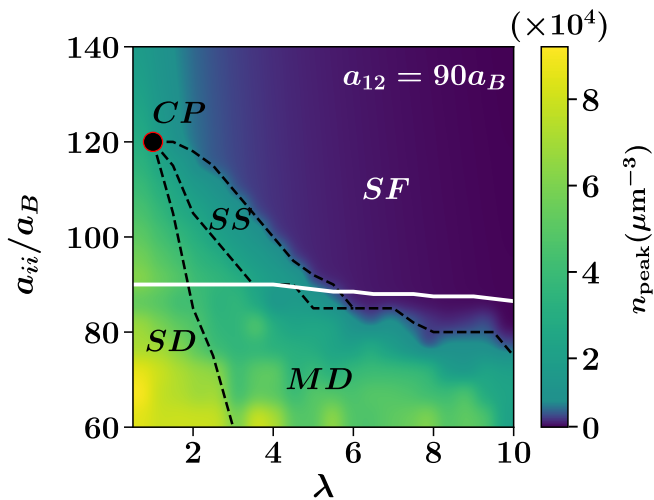}
	\caption{Shows the groundstate phase diagram of a Dy-Dy mixture as a function of intra-component s-wave scattering length ($a_{ii}$) and trap aspect ratio ($\lambda$). The shading represents the peak density of the combined binary system. The black dashed lines indicate phase boundaries between the superfluid state (SF) with a low-peak density, a supersolid (SS) and a droplet state (SD and MD) with a high peak density. The large black dot encircled by a red border demarks the critical point (CP) and the white solid line has been drawn at $\Lambda=0.5$, representing the immiscibility phase boundary. Results are for the case of $N_1=N_2=6\times 10^4$, $^{164}$Dy atoms with $a_{ii}^{dd}=131a_B~(\mu_i^{m}=9.93\mu_B)$ where $a_B$ is the Bohr radius and $\mu_B$ is Bohr magneton.}\label{fig:2}
\end{figure}
\subsubsection{Intra-species scattering length vs. trap aspect ratio ($\lambda$)}
So far, we have discussed the effect of intra- and inter-species contact interactions on the groundstate of a binary dipolar mixture for different numbers of atoms. However, the trap aspect ratio $\lambda=\omega_z/\omega$ (trap geometry) is also one of the key parameters to explore different possible groundstate phases. Trap geometry influences the condensate shape as well as the DDI energy. The average DDI energy changes from negative to positive as the shape of the condensate changes from prolate to oblate. To construct a phase diagram with $\lambda$ and intra-species scattering length $a_{ii}(i=1,2)$, we fix the inter-species scattering length at $a_{12}=90a_B$ and number of particles at $N_1=N_2=6\times 10^4$. In Fig. \ref{fig:2}, we plot the peak density corresponding to the groundstate of a binary mixture as a function of $\lambda$ and $a_{ii}$. The peak density results emphasize a significant change in the density among the SF and SS, droplet (SD and MD) phases  (The SS and droplet phases are approximately two orders of magnitude denser than the SF phase). We demark all these phase boundaries by black dashed lines. All these phase transition lines terminate at a critical point (CP). Beyond this critical point, there is no abrupt phase transition. Rather a smooth evolution among the above-mentioned phases is observed. A similar kind of behavior was also observed for a single component dBEC \cite{bisset_2016_groundstate}. Here, the immiscibility boundary is close to $a_{12}=\sqrt{a_{11}a_{22}}=90a_B$ (as we discussed earlier for a large number of atoms) marked by the white solid line drawn at $\Lambda=0.5$ in Fig. \ref{fig:2}. The region below ($a_{ii}<a_{12}$) and above $(a_{ii}>a_{12})$ the white solid line corresponds to the phase-separated (immiscible) and miscible phase domains, respectively.

\begin{figure}[tb!]
	\centering
	\includegraphics[width=0.485\textwidth]{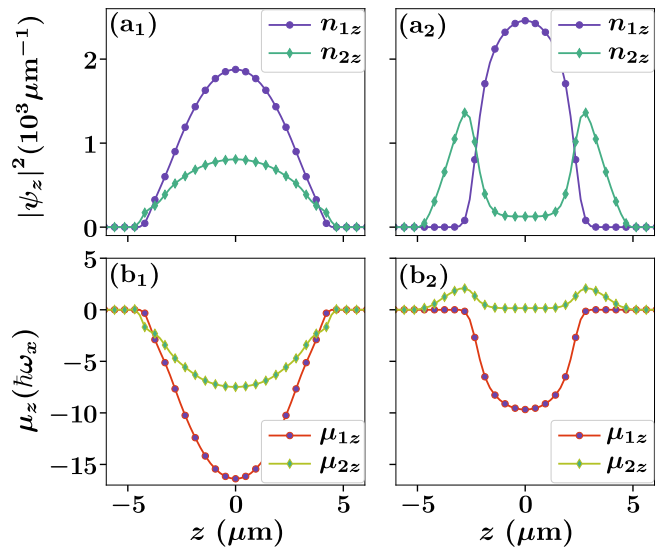}
	\caption{Miscible and immiscible SD state in an imbalanced mixture. $\rm(a_1),~(a_2)$ Shows the integrated axial density of species-1 (violet) and species-2 (green) of a miscible and immiscible SD state, respectively. $\rm(b_1),~(b_2)$ Show the corresponding chemical potential densities along the axial direction (z-axis). Results are for the case of $a_{11}=70a_B,~a_{22}=80a_B,~N_1=10^4,~N_2=5\times 10^3$ and $(\rm a_1),(b_1)$ $a_{12}=65a_B~(<\sqrt{a_{11}a_{22}})$, $(\rm a_2),(b_2)$ $a_{12}=75a_B~(>\sqrt{a_{11}a_{22}})$. In both cases the binary mixture is confined in an oblate shape harmonic trap with $(\omega_x,\omega_y,\omega_z)=2\pi\times(45,45,133)$Hz.}\label{fig:1D_chpt_den_imbalanced_sd}
\end{figure}

\section{Supersolid and droplets state in an imbalanced mixture}\label{seciv}
Now we consider an imbalanced binary mixture, where the intra-species interactions and number of particles among the components are not equal $(a_{11}\neq a_{22}$ and $N_1\neq N_2$). In addition to all the possible groundstates discussed so far, some mixed states like a mixture of SS-SF, SS-MD, and SS-SS states are formed in this case. Here, we consider a Dy-Dy mixture with intra-species scattering lengths $a_{11}=70a_B$, $a_{22}=80a_B$, and the condensates contain $N_1=10^4$ and $N_2=5\times 10^3$ number of atoms. With these chosen values of parameters, the binary mixture undergoes a miscible SD to immiscible SD phase transition beyond $a_{12}\approx 75a_B$. To look into these miscible and immiscible SD states of the imbalanced mixture, we depict the integrated density profiles $n_{1z}$ and $n_{2z}$ of both species along the axial direction in Figs. \ref{fig:1D_chpt_den_imbalanced_sd}$(\rm a_1)$ and \ref{fig:1D_chpt_den_imbalanced_sd}$(\rm a_{2})$, respectively. In the first scenario with $a_{12}=65a_B$ ($< \sqrt{a_{11}a_{22}}$), the density profiles of both the species completely overlap with each other and form a miscible SD state (see Fig. \ref{fig:1D_chpt_den_imbalanced_sd}$(\rm a_1))$. However, as we increase $a_{12}$ beyond the miscible to immiscible transition value, species-1 (major component) remains at the center, due to its larger population (atom number) and smaller intra-species scattering length. The species-2 (minor component) is pushed along the axial direction and resides at each extreme end of the domain formed by the major component (see Fig. \ref{fig:1D_chpt_den_imbalanced_sd}$(\rm a_2))$. See Appendix \ref{D} for the discussion on the effective potential experienced by each species due to the presence of the other component.
\begin{figure}[tb!]
	\centering
	\includegraphics[width=0.48\textwidth]{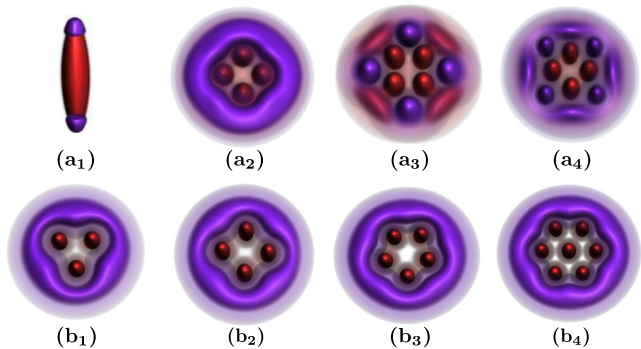}
	\caption{Shows the 3D isosurface density profiles of (species-1 (red), species-2 (violet)) immiscible groundstates in an imbalanced binary mixture, drawn at $10\%$ of the respective peak densities. ($\rm a_1$) Immiscible SD state along the $z$ direction, parameters remain same as of Fig. \ref{fig:1D_chpt_den_imbalanced_sd}($\rm a_2$). $\rm (a_2){-}(a_4$) show various mixed ground states like ($\rm a_2$) SS-SF, ($\rm a_3$) SS-MD, and ($\rm a_4$) SS-SS states in $x-y$ plane. $\rm (b_1){-}(b_4)$ show different pattern formation at the interface of two species in an imbalanced mixture. Results are for the case of a binary dBEC confined in a harmonic trap with $\{\omega_{x},\omega_y,\omega_z\}=2\pi\times\{45,45,133\}$ Hz, $\rm (a_2){-}(a_4)$ $a_{11}=90a_B,~ a_{22}=95a_B,~a_{12}=95a_B,$ and $N_1=6\times 10^4$, and for $\rm (b_1){-}(b_4)$ $a_{11}=90a_B,~ a_{22}=100a_B,~a_{12}=100a_B$. Other parameters are $(\rm a_2)$ $N_2=5\times10^3$, $(\rm a_3)$ $N_2=2\times10^4$, $(\rm a_4)$ $N_2=5\times10^4$ $(\rm b_1)$ $N_1=6\times10^4,~N_2=10^3$, $(\rm b_2)$ $N_1=6\times10^4,~N_2=5\times10^3$, $(\rm b_3)$ $N_1=8\times10^4,~N_2=1.5\times10^4,$ and $(\rm b_4)$ $N_1=10^5,~N_2=3\times10^4.$}\label{fig:SS_SF_Pattern}
\end{figure}

The reason behind these kinds of density distributions can be clearly understood from the chemical potential densities along the axial direction ($\mu_{z}$) of each species as shown in Figs. \ref{fig:1D_chpt_den_imbalanced_sd}($\rm b_1,~b_2$). In the miscible SD state, the chemical potential densities of each component are negative indicating that both components are self-bound. Despite having a different number of particles and intra-species scattering lengths, the large negative chemical potential of the major component sets the spatial width of both species equal (see Fig. \ref{fig:1D_chpt_den_imbalanced_sd}($\rm b_1$)). The chemical potential of each species increases with $a_{12}$. In the immiscible SD and MD regimes, the chemical potential density of the minor component becomes positive. However, due to the negative chemical potential density of the major component along the axial direction, the minor component is bound at each end of the domain formed by the major component (Fig. \ref{fig:1D_chpt_den_imbalanced_sd}($\rm b_2$)). We have shown the corresponding 3D isosurface density profile of the immiscible SD state in Fig. \ref{fig:SS_SF_Pattern}($\rm a_1$). In absence of the major component, the minor component can not bind itself in these axial positions. The total chemical potential of the binary mixture in this state is still negative, which 
{\unskip\parfillskip 0pt\par}
\begin{figure}[h]
	\centering
	\includegraphics[width=0.47\textwidth]{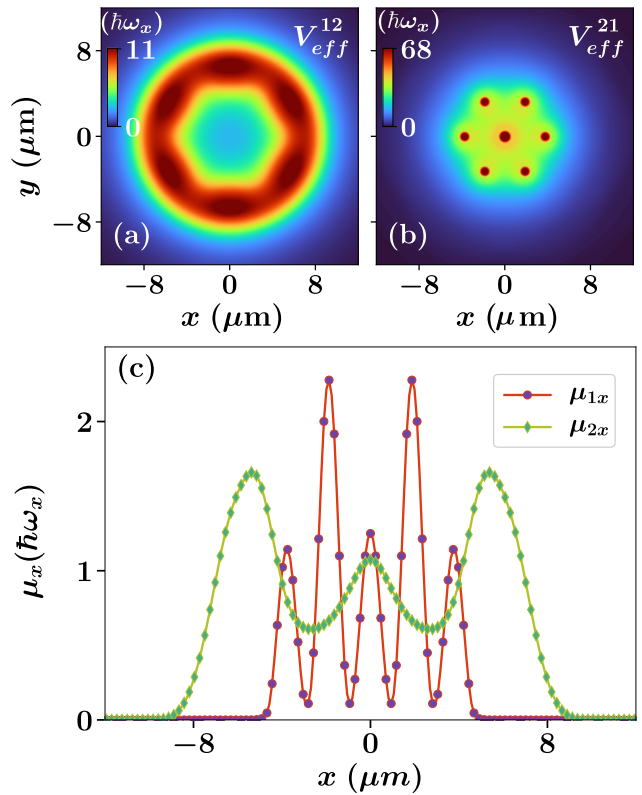}
	\caption{(a)-(b) Shows the effective potential due to the presence of the other component of a SS-SF mixed state in the x-y plane. Colorbar represents the effective potential in units of $\hbar \omega_x$. (c) Shows the corresponding chemical potential densities of species-1 (red), and species-2 (light-green) along the x-axis. Results are for the case of $a_{11}=90a_B,~a_{22}=100a_B,~a_{12}=100a_B,~N_1=10^5,~N_2=3\times 10^4$, and the imbalanced binary mixture is confined in a harmonic trap potential with $(\omega_x,\omega_y,\omega_z)=2\pi\times(45,45,133)$Hz. }\label{fig:2D_imbalance_mix_chem_veff}
\end{figure}
\noindent implies that together they form a self-bound immiscible droplet state. \par  
Mixed groundstates can be formed when both the condensates of the binary mixture have comparatively large intra-species scattering lengths and form partially or completely phase-separated (immiscible) groundstates. Various groundstates of mixed phases like SS-SF, SS-MD and SS-SS formed in a binary dBEC depending upon the number of atoms, intra- and inter-species scattering lengths. The 3D isosurface density profiles of these mixed states are shown in Fig. \ref{fig:SS_SF_Pattern}($\rm a_2-a_4$). In this regime, beyond a critical value of $a_{11}$ and $a_{22}$ (here we consider $a_{11}<a_{22}$), both the components have a slightly positive chemical potential. Further, the first species with smaller intra-species interaction and larger number of atoms occupies the central position of the trap, similar to the previous case. However, due to the positive chemical potential, it (the major component) can not hold the second species at each end along the axial direction. Rather in the presence of a harmonic trap, the minor 
{\unskip\parfillskip 0pt\par}
\onecolumngrid

\begin{figure}[H]
	\centering
	\includegraphics[width=\textwidth]{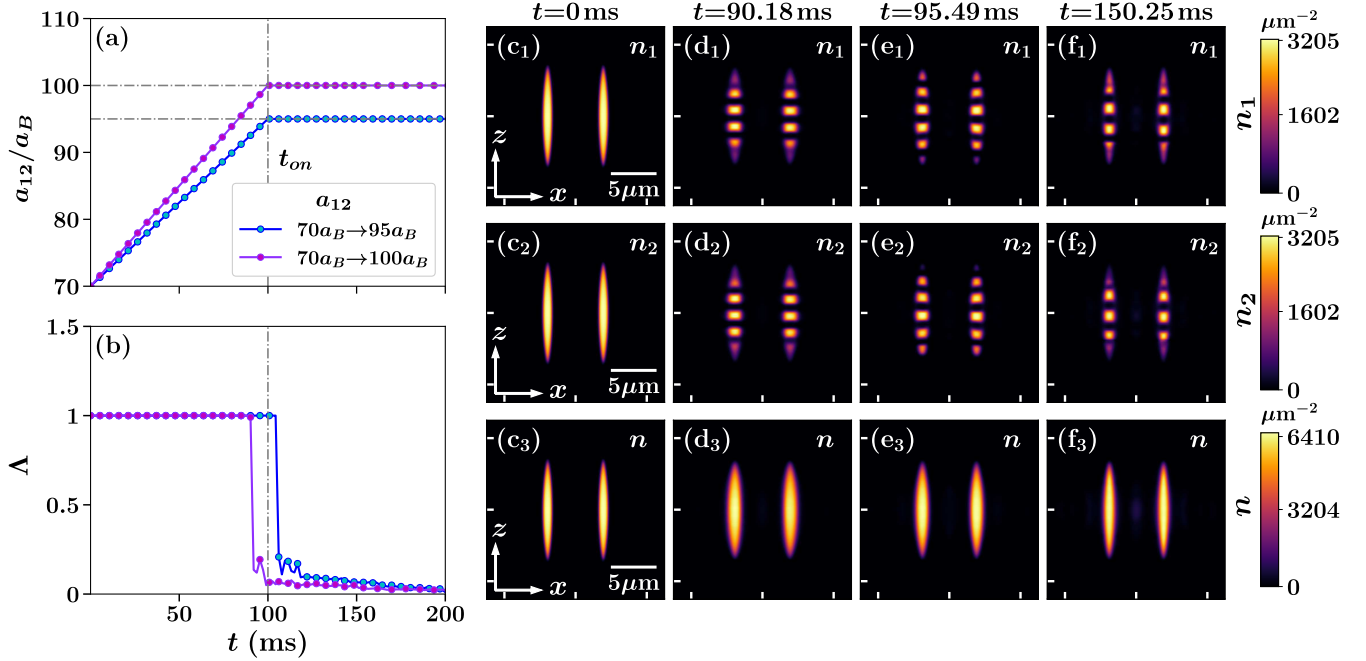}
	\caption{Tuning the inter-species scattering lengths ($a_{12}$) for the fixed values of $a_{ii}=90a_B$. Panel (a) shows the linear ramp with ramp time $t_{on}=100$ms. From an initial value of $a_{12}=70a_B$, the inter-species scattering lengths are tuned to the final values $a_{12}=95a_B$ and $a_{12}=100a_B$, respectively, through this ramp. Panel (b) shows the corresponding variation of the overlap integral with time. ($\rm c_1-f_3$) Shows the snapshots of the density profile of species-1 $n_1(x,z,y=0)$ ($\rm c_1-f_1$), species-2 $n_2(x,z,y=0)$ ($\rm c_2-f_2$), and of the total system $n(x,z,y=0)$ ($\rm c_3-f_3$), following a quench from a miscible MD state with $a_{12}=70a_B$ to an immiscible MD state with $a_{12}=100a_B$. Both the condensate consist of $N=6\times 10^4$ atoms, and the balanced binary mixture is confined in a harmonic trap with $(\omega_x,\omega_y,\omega_z)=2\pi\times(45,45,133)$Hz. The colorbar denotes the density in units of $\mu \rm m^{-2}$. }\label{fig:Dynamics_quench_aij}
\end{figure}
\twocolumngrid
\noindent component is pushed along the radially outward direction as illustrated in Figs. \ref{fig:SS_SF_Pattern}$(\rm (a_2){-}(a_4))$ for the scattering lengths $a_{11}=90a_B$, $a_{22}=95a_B$ and $a_{12}=95a_B$ and the species-1 contains $N_1=6\times 10^4$ number of atoms. For different numbers of atoms in the second species $N_2=10^3,~2\times 10^4$ and $5\times 10^4$ we observe different mixed phases like SS-SF (Fig. \ref{fig:SS_SF_Pattern}$(\rm a_2)$), SS-MD (Fig. \ref{fig:SS_SF_Pattern}$(\rm a_3)$) and SS-SS (Fig. \ref{fig:SS_SF_Pattern}$(\rm a_4)$), respectively. The effective potential experienced by each species due to the presence of the other species plays a crucial role in determining the position of the condensates in the trap. In Figs. \ref{fig:2D_imbalance_mix_chem_veff}(a), \ref{fig:2D_imbalance_mix_chem_veff}(b), we have shown the effective potential experienced by each species in the $x-y$ plane for a SS-SF mixed state corresponding to the density profile as shown in Fig. \ref{fig:SS_SF_Pattern}($\rm b_4$). The corresponding chemical potential densities along the $x$-axis ($\mu_x$) are also shown in Fig. \ref{fig:2D_imbalance_mix_chem_veff}(c). As we explained earlier, both the condensates have positive chemical potential densities along the $x$-axis (see Fig. \ref{fig:2D_imbalance_mix_chem_veff}(c)). Moreover, the first species experiences a minimum effective potential at the trap center while the second species finds the same at the periphery of the first condensate and forms a radially immiscible mixture.\par
Interestingly enough, in an SS-SF mixed state, various polygonal shape patterns form at the interface of the two species depending upon the number of droplets in the SS state, as shown in Figs. \ref{fig:SS_SF_Pattern}$\rm ((b_1){-}(b_4$)). The number of droplets can be varied by changing either the number of atoms or the intra-species scattering lengths. For the visualization of these polygonal patterns, we choose the intra-, and inter-species scattering lengths to be $a_{11}=90a_B,~a_{22}=100a_B$ and $a_{12}=100a_B$, and the number of atoms $(N_1,N_2)$ to be $(6\times 10^4,~10^3),~(6\times10^4,~5\times10^3),~(8\times10^4,~1.5\times 10^4),~\rm{and}~(10^5,3\times10^4)$ which are corresponding to the triangular (Fig. \ref{fig:SS_SF_Pattern}($\rm b_1$)), rectangular (Fig. \ref{fig:SS_SF_Pattern}($\rm b_2$)), pentagonal (Fig. \ref{fig:SS_SF_Pattern}($\rm b_3$)) and hexagonal (Fig. \ref{fig:SS_SF_Pattern}($\rm b_4$)) shapes patterns, respectively, at the interface.\par 

\section{Dynamics}\label{secv}
So far we have discussed the phase diagrams of a balanced binary mixture and different possible groundstates in an imbalanced binary mixture. Now we explore the effect of tuning intra- and inter-species scattering lengths of a balanced binary mixture in real-time dynamics. Consider the first case where we initially prepare the dBEC in a miscible MD regime, with $a_{ii}=90a_B,~ a_{12}=70a_{B}~ \rm{and}~ N_1=N_2=6\times 10^4$. We then perform two different slow linear ramps for increasing the value of $a_{12}$, one from $70a_B$ to $95a_B$, and the other from $70a_B$ to $100a_B$ over a ramp time $t_{on}=100$ ms. After that, the inter-species scattering length $a_{12}$ is kept constant to check the stability of the evolved system (see Fig. \ref{fig:Dynamics_quench_aij}(a)). We find both the evolutions produce dynamically stable droplets, and these results are also consistent with the formation of a self-bound droplet state in a trap-less system \cite{smith_2021_quantum,smith_2021_approximate}. In Fig. \ref{fig:Dynamics_quench_aij}(b) and \ref{fig:Dynamics_quench_aij}(c), we have shown the time evolution of the overlap integral ($\Lambda$) and the density profile of each species in the $x-z$-plane ($n_i(x,z,y=0)$), respectively.\par
\begin{figure}[tb!]
	\centering
	\includegraphics[width=0.47\textwidth]{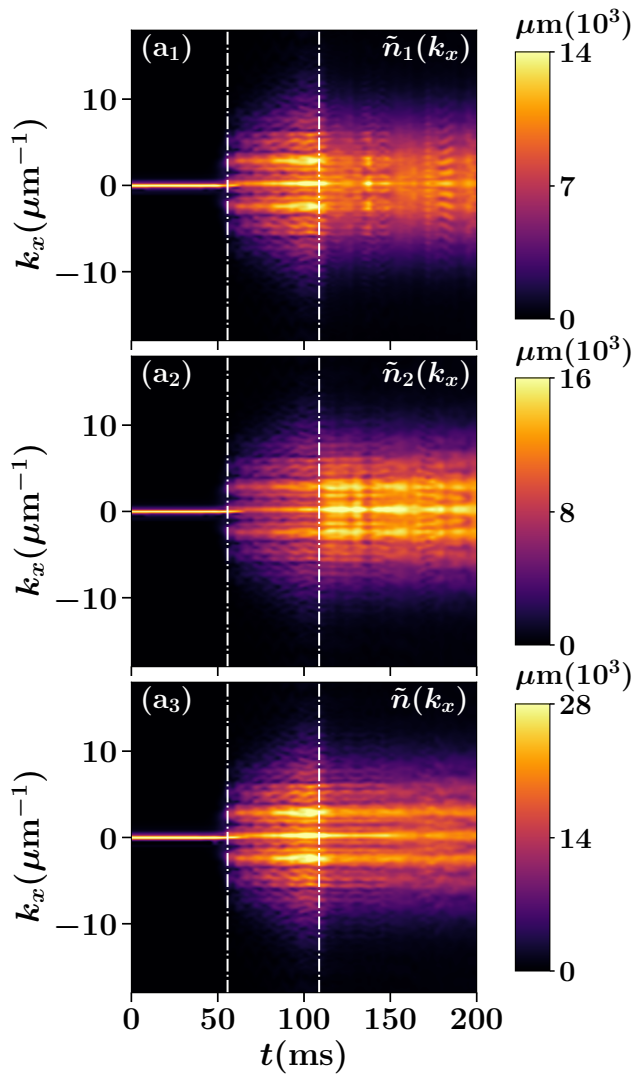}
	\caption{Shows the time evolution of the momentum space density ($\rm a_1$) of species-1 $\tilde{n}_1(k_x)$, ($\rm a_2$) species-2 $\tilde{n}_2(k_x)$, and ($\rm a_3$) of the total system $\tilde{n}(k_x)$, following a linear quench with ramp time $t_{on}=100$ms from a miscible SF state with $a_{ii}=120a_B$ to a immiscible MD state with $a_{ii}=85a_B$. Both the condensate consists of $N=6\times 10^4$ atoms, $a_{12}=90a_B$, and the balanced binary mixture is confined in a harmonic trap with $(\omega_x,\omega_y,\omega_z)=2\pi\times(45,45,133)$Hz. The colorbar denotes the density in the $k_x$-space, in units of $10^3\mu \rm m$.}\label{fig:Dyna_FFT_Quench_aii_120_85_t}
\end{figure}
Initially, while $a_{12}<a_{ii}$, the mixture forms a miscible MD state. As soon as $a_{12}>a_{ii}$, the system undergoes a miscible to immiscible transition. Near the transition time, the value of the overlap integral rapidly changes from 1 to 0. Due to this sudden change, each component forms multiple periodic segregated domains along the axial direction and forms a completely phase-separated density profile. In this state, the density profile of each component is complementary to the other, and together they form an axially symmetric immiscible MD state.\par
\begin{figure}[tb!]
	\centering
	\includegraphics[width=0.48\textwidth]{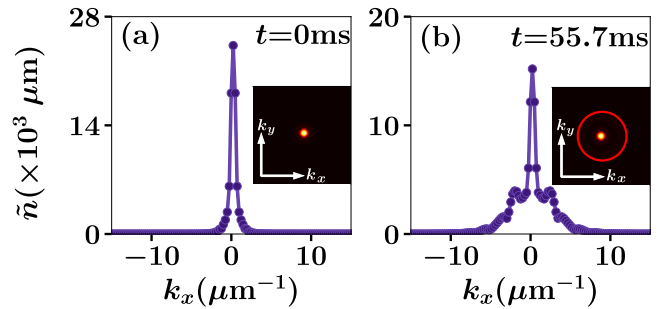}
	\caption{Snapshots of the density profile of the composite binary system in momentum space following a linear quench with ramp time $t_{on}=100$ms from a miscible SF state with $a_{ii}=120a_B$ to an immiscible MD state with $a_{ii}=85a_B$ keeping the inter-species scattering length $a_{12}$ fixed at $90a_B$. (a) Initially at $t=0$ms, a single peak corresponds to the SF state at $k_x=0~\mu \rm m^{-1}$, (b) at $t=55.7$ ms,appearance of two symmetric roton peak at $k_x=\pm 3.05~\mu \rm m^{-1}$. The insets show the corresponding density distribution $\tilde{n}(k_x,k_y,k_z{=}0)$ in ($k_x,k_y$)-plane. Parameters are same as of Fig. \ref{fig:Dyna_FFT_Quench_aii_120_85_t}.
	}\label{fig:Dyna_FFT_Quench_aii_120_85}
\end{figure}
Subsequently, we also explore the quench dynamics of a balanced binary mixture, starting from a SF state in a miscible regime. In this case, dynamics are triggered by reducing the intra-species scattering lengths into a miscible and immiscible SS regime. For a fixed inter-species scattering length at $a_{12}=90a_B$, we perform two interaction quenches by linearly reducing the intra-species scattering length $a_{ii}$, one from $120a_B$ to $95a_B $, and the other from $120a_B$ to $85a_B$ over a time period $t_{on}=100$ ms, after which $a_{ii}$ is held constant (see Fig. \ref{fig:Dynamics_quench_aii}(a) in Appendix \ref{E}), and we observe the time evolution of the binary system. In both quenching processes, as $a_{ii}$ is reduced, the system undergoes a roton instability at $a_{ii}\approx 100a_B$. In Fig. \ref{fig:Dyna_FFT_Quench_aii_120_85_t}, we have shown the time evolution of momentum space density $\tilde{n}(k_x)$, following the quench $a_{ii}=120a_B$ to $85a_B$. Initially, up to $t=55.7$ ms, the binary mixture forms a miscible SF state which corresponds to a single density peak at $(k_x,k_y)=(0,0)~\mu \rm m^{-1}$ (see Fig. \ref{fig:Dyna_FFT_Quench_aii_120_85}(a)). Later following the roton\footnote{The roton modes are characterized by the quantum number $m$ \cite{schmidt_2021_roton,hertkorn_2021_supersolidity,chomaz_2018_observation}. In the $(k_x-k_y)$ plane, the roton population is spread over a ring which corresponds to a radial roton mode with $m=0$.} instability at the phase boundary, a ring of radius $3.05~\mu \rm m^{-1}$ is readily visible in the $k_x{-}k_y$ plane and for $k_y=0$, the density profile $n(k_x,k_y=0)$ corresponds to the appearance of two additional side peaks in the momentum space (similar like a cigar-shaped trap geometry) (see Fig. \ref{fig:Dyna_FFT_Quench_aii_120_85}(b)). The symmetric side peaks in the momentum space essentially indicate a periodic density modulation in the real space. The binary dipolar mixture forms a miscible SS state in the time interval $t=55.7$ ms ($a_{ii}=100.5a_B$) and $t=85.7$ ms ($a_{ii}=90a_B$). As the $a_{ii}$ is reduced further the system enters into a MD phase domain and when $a_{ii}<a_{12}$, the overlap integral ($\Lambda$) changes from 1 to 0 rapidly (see Fig. \ref{fig:Dynamics_quench_aii}(b)), and the system forms an immiscible MD state. The characteristic density snapshots, while performing the quench of $a_{ii}$ from $120a_B$ to $95a_B$, in the x-y plane are presented in Figs. \ref{fig:Dynamics_quench_aii}($\rm (c_1)-(e_3)$ in Appendix \ref{E}).


\section{Conclusions and outlook}\label{secvi}
In this work, we have theoretically investigated the scope of formation of two-dimensional supersolid and droplet lattice states in a binary dBEC. We performed an in-depth investigation and demonstrated that a binary dipolar mixture confined in a circular symmetric trap could exhibit a large variety of groundstate phases with rich properties inaccessible for a non-dipolar binary mixture and in a single component dBEC. The emergent phases include SF, SS, SD, and MD states both in miscible and immiscible phase domains. The interplay between intra-, inter-species contact interaction and the anisotropic dipole-dipole interaction leads to the formation of all these phases. Numerically solving the coupled eGPE, we obtain all these results. Besides the 3D numerical simulation, we also employ a variational approach in the SSA framework to validate our results. Although in this work we have demonstrated the results for a Dy-Dy mixture and a specific range of atom-atom interaction strengths, our analysis can be considered as one step forward in the direction of the formation of more exciting new phases in binary dipolar BECs yet to be revealed in ongoing and future research works.\par
We have examined different groundstate phases in a balanced binary dBEC, and depicted the phase diagrams as a function of the number of particles, intra-, and inter-species scattering lengths. We also monitor the effect of trap geometry in terms of trap aspect ratio on the groundstate phases. More intriguing mixed phases appear for an imbalanced mixture. In the miscible phase domain, both condensates possess exactly identical shapes. Whereas in the immiscible phase domain, we observe two types of immiscible phases: (i) axially immiscible phase, and (ii) radially immiscible phase. The axially immiscible phase for a self-bound droplet state without any trapping confinement is predicted in some of the recent theoretical works. However, the radially immiscible phases have not been reported so far to the best of our knowledge.\par
In the self-bound immiscible droplet regime, due to the dominant anisotropic dipole-dipole interaction, the component with a larger atom number and smaller intra-species scattering length takes the central position of the droplet and forms two potential minima at its two outer edges along the axial direction. The second component with a slightly positive chemical potential energy is docked at the above-mentioned position by the first component with a negative chemical potential energy and forms an axially immiscible self-bound droplet state. The chemical potential of each condensate increases with the increase of intra- and inter-species scattering length. Hence in an immiscible regime for a comparatively large value of intra-species scattering length, the chemical potential of both the component becomes positive, and no longer the major component can hold the minor component at the axial position. For an imbalanced mixture in the presence of a circularly symmetric harmonic trap, the minor component with a comparatively small number of atoms and large intra-species scattering length is pushed along the radially outward direction and forms a radially immiscible state. Depending on the value of intra-species scattering lengths, each species can form a MD, SS, or SF state. Whereas for a balanced system, none of the condensates have such biasness due to equal interaction strength and hence forming a radially asymmetric phase-separated state.\par
Utilizing our groundstate phase diagrams for a balanced binary mixture as a reference, we explore the dynamics across the phase boundaries by tuning the interaction strengths. The dynamical transition across the phase boundaries initially governs some instability in the system, leading to the formation of some metastable states in the intermediate time scale. In long-time dynamics, we have shown the dynamical phase transition from a miscible droplet state to an immiscible droplet state with multiple domains and a crossover from a SF state to a MD droplet state via a SS state. \par
Our observations pave the way for several future research directions. In this work, we have restricted our study to a particular Dy-Dy (homonuclear) mixture. However, it would be intriguing to explore the formation of different possible phases in a heteronuclear binary dipolar mixture like Er-Dy mixture \cite{trautmann_2018_dipolar,politi_2022_interspecies,durastante_2020_feshbach}. Furthermore, one straightforward option is to investigate the lifetime of these phases by incorporating the effect of three-body interaction loss \cite{chomaz_2019_longlived,bottcher_2019_transient}. Another intriguing direction would be to consider the impact of thermal fluctuation and unravel corresponding phases as well as dynamical nucleation of the supersolid and droplet lattice in the finite temperature limit \cite{sanchez-baena_2022_heating,derosi_2021_thermal}. Moreover, the evaporation cooling mechanism is an alternative approach to the interaction quench and provides the prospects of forming a long-lived 2D supersolid state in a binary dipolar mixture \cite{bland_2022_twodimensional}. Another vital prospect would be to investigate quantum turbulence \cite{johnstone_2019_evolution,gauthier_2019_giant,das_2022_vortex}, pattern formation \cite{hertkorn_2021_pattern,zhang_2021_phases,maity_2020_parametrically,kwon_2021_spontaneous} and various topological excitations such as the formation of vortex clusters and solitary waves in a binary dipolar condensate.  Finally, the observation discussed in this work would be equally fascinating beyond the Lee-Huang-Yang description \cite{ota_2020_leehuangyang,hu_2020_consistent}.

\begin{acknowledgments}
We thank Koushik Mukherjee and S.I.Mistakidis for fruitful discussions. We also acknowledge the National Supercomputing Mission (NSM) for providing computing resources of ‘PARAM Shakti’ at IIT Kharagpur, which is implemented by C-DAC and supported by the Ministry of Electronics and Information Technology (MeitY) and Department of Science and Technology (DST), Government of India.
\end{acknowledgments}

\appendix
\section{Numerical Methods}\label{A}
Results in this work are based on three-dimensional numerical simulations in the coupled eGPE (Eq. (\ref{egpe})) framework. For the sake of the convenience of numerical simulations and better computational precision, we cast the coupled eGPE into a dimensionless form. This is achieved by rescaling the length scale and time scale in terms of oscillator length $l_{osc}=\sqrt{\hbar/m\omega_x}$ and $\omega_x$ trapping frequency along the x direction. Under this transformation, the wave function of species-$i$ obeys $\psi_i(\vb{r}^{\prime})=\sqrt{l_{osc}^3/N_i}\psi_i(\vb{r})$, where $N_i$ is the number of particles in species-$i$. After the transformation of variables into dimensionless quantities the coupled eGPE is solved by split-step-Crank-Nicolson scheme \cite{crank_nicolson_1947}. Since the dipolar potential has a singularity at $r=0$ (see Eq. (\ref{dip_pot})), it is numerically evaluated in Fourier space and we obtain the real space contribution through the application of the convolution theorem. The groundstates of binary dipolar condensate are obtained by propagating the relevant equations in imaginary time until the relative deviations of the wave functions (calculated at every grid point) and energy of each condensate between successive time steps are less than $10^{-6}$ and $10^{-7}$, respectively. Furthermore, we fix the normalization of each species at every time instant of the imaginary time propagation. Using this groundstate solution as an initial state, at $t=0$, and by changing the interaction strengths we monitor their evolution in real-time. Our simulations are performed within a 3D box grid containing $(256\times 256\times 256)$ grid points, with the spatial grid spacing $\Delta_x=\Delta_y=\Delta_z=0.1 l_{osc}$ while the time step $\Delta_t=10^{-4}/\omega_x$.

\begin{figure}[tb!]
	\centering
	\includegraphics[width=0.48\textwidth]{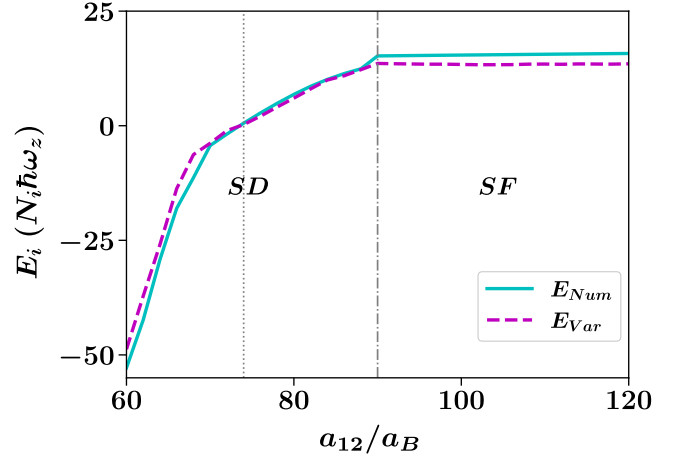}
	\caption{Shows the variation of energy with the inter-species scattering length $a_{12}$ of a balanced binary mixture. The mixture is confined in a harmonic trap with $(\omega_x,\omega_y,\omega_z)=2\pi\times(45,45,133)$Hz and the other parameters fixed at $a_{11}=a_{22}=90a_B$ and $N_1=N_2=10^4$ number of atoms. The results are obtained through a variational approach within the SSA framework (dashed line), and the coupled eGPE solutions (solid line) are in agreement.}\label{fig:Binary_BEC_energy_var_num_10000_a11_90}
\end{figure}
\section{Variational solution within same shape approximation (SSA) framework}\label{B}
In addition to numerical 3D simulations of Eq. (\ref{egpe}), we employ a simple variational approach in the regime where both components are miscible and take the exact same shape (i.e., $\psi_1=\psi_2$), and this is only possible when both the condensate have an equal number of atoms and equal intra-species interaction. In this regime, the Hamiltonian of the $i$'th condensate is reduced to an effective single-species Hamiltonian given by
\begin{eqnarray}
	&&\hat{\mathcal{H}}_{SSA}^{i}=-\frac{\hbar^2}{2m_i}\nabla^2+V_t(\textbf{r})+\mathcal{G}_{SSA}\abs{\psi_i(\textbf{r})}^2+\nonumber\\&&\mathcal{G}_{SSA}^{dd}\int d\textbf{r}^{\prime}U^{dd}(\textbf{r}-\textbf{r}^{\prime})\abs{\psi_i(\textbf{r}^{\prime})}^2+\gamma_{SSA}^{QF}(\epsilon_{ii}^{dd})\abs{\psi_i(\textbf{r})}^3.\label{hami}\nonumber\\
\end{eqnarray}
Here $\mathcal{G}_{SSA}=g_{ii}+g_{12}$, $\mathcal{G}_{SSA}^{dd}=g_{ii}^{dd}+g_{12}^{dd}$ are the effective strengths of contact interaction and DDI, respectively. The last term of Eq. \ref{hami} denotes the contribution of quantum fluctuation. We remark that within this SSA framework, quantum fluctuations depend on the density $n_i^{3/2}$, where $n_i=\abs{\psi_i}^2$. The coefficient of quantum fluctuations $\gamma_{SSA}^{QF}$ is well approximated by the known form of a single-species dBEC \cite{lima_2011_quantum}:
\begin{eqnarray}
	\gamma_{SSA}^{QF}(\epsilon_{SSA}^{dd})=&&\frac{32}{3}\mathcal{G}_{SSA}\sqrt{\frac{a_{SSA}^3}{\pi}}\nonumber\\&&\Re{\int_{0}^{1}\dd u[1+\epsilon_{SSA}^{dd}(3u^2-1)]^{5/2}},\nonumber\\
\end{eqnarray}
where $a_{SSA}=a_{ii}+a_{12}$ and the dimensionless parameter $\epsilon_{SSA}^{dd}=\mathcal{G}_{SSA}^{dd}/\mathcal{G}_{SSA}$ quantifies the effective relative strength of the DDI to the contact interaction.
Within this SSA framework, the total energy of the $i$'th species
\begin{eqnarray}
	&&E_i=\int \bigg[\frac{\hbar^2}{2m_i}\abs{\nabla\psi_i}^2+v_t(\textbf{r})\abs{\psi_i(\textbf{r})}^2+\frac{\mathcal{G}_{SSA}}{2}\abs{\psi_i(\textbf{r})}^4\nonumber\\&&+\frac{\mathcal{G}_{SSA}^{dd}}{2}\int d\textbf{r}^{\prime}U^{dd}(\textbf{r}-\textbf{r}^{\prime})\abs{\psi_i(\textbf{r}^{\prime})}^4+\frac{2}{5}\gamma_{SSA}^{QF}\abs{\psi_i(\textbf{r})}^5\bigg]d\textbf{r}.\label{energy}\nonumber\\
\end{eqnarray}
\begin{figure}[tb!]
	\centering
	\includegraphics[width=0.48\textwidth]{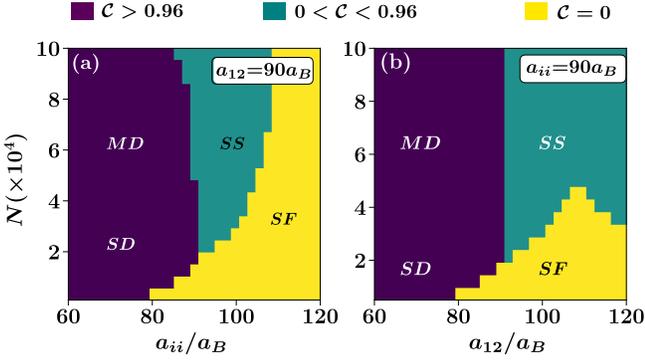}
	\caption{Groundstate phase diagrams in an oblate harmonic trap of a Dy-Dy mixture as a function of number of particles in each condensate and (a) intra-, (b) inter-species scattering length. The color domains on the phase plots correspond to the different contrast regimes as demarked above in the phase diagrams. Other parameters remain same as of Fig. \ref{fig:1}.}\label{fig:contrast_phase_diagram}
\end{figure}
A qualitative and to some extent quantitative insight into the droplet and supersolid physics in the miscible SSA regime may be gained from a simplified Gaussian ansatz
\begin{equation}
	\psi_i(\textbf{r})=\sqrt{\frac{8N_i}{\pi^{3/2}\sigma_{\rho}^2\sigma_z}}\prod_{\eta=\rho,z}\exp(-\frac{2\eta^2}{\sigma_{\eta}^2}),\label{var_psi}
\end{equation}
where the variational parameters are the condensate widths $\sigma_{\eta}$ in the $\eta=\rho,z$ direction. We insert the ansatz (\ref{var_psi}) into Eq. (\ref{energy}) and obtain
\begin{eqnarray}
	\frac{E_i}{N_i\hbar\omega_z}&=& \left(\frac{2 l_z^2}{\sigma_{\rho}^2}+\frac{l_z^2}{\sigma_z^2}\right)+\left(\frac{1}{8\lambda^2}\frac{\sigma_{\rho}^2}{l_z^2}+\frac{\sigma_z^2}{16l_z^2}\right)\nonumber\\&&+\frac{4N_i\left(\mathcal{G}_{SSA}-\mathcal{G}_{SSA}^{dd}f(\frac{\sigma_{\rho}}{\sigma_z})\right)}{(2\pi)^{3/2}\sigma_{\rho}^2\sigma_z\hbar\omega_z}\nonumber\\&&+\frac{128N_i^{3/2}\gamma_{SSA}^{QF}}{25\sqrt{5}\pi^{9/4}\sigma_{\rho}^3\sigma_z^{3/2}\hbar\omega_z},\label{var_energy}
\end{eqnarray}
where, 
\begin{equation}
	f(k)=\frac{1+2k^2}{1-k^2}-\frac{3k^2\tanh^{-1}{\sqrt{1-k^2}}}{(1-k^2)^{3/2}},
\end{equation}
{\unskip\parfillskip 0pt\par}
\begin{figure}[tb!]
	\centering
	\includegraphics[width=0.48\textwidth]{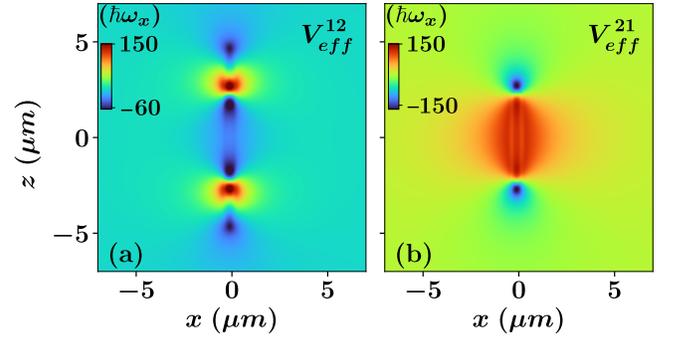}
	\caption{Shows the effective potential due to the presence of the other component of an immiscible SD state in the x-z plane. Colorbar represents the effective potential in units of $\hbar \omega_x$. Other parameters remain same as of Fig. \ref{fig:1D_chpt_den_imbalanced_sd}($\rm a_2$). }\label{fig:2D_veff_immis_sd}
\end{figure}
\noindent and $l_z=\sqrt{\hbar/m_i\omega_z}$. We find the stationary solutions by numerically locating the values of $\sigma_{\rho}$ and $\sigma_z$ that minimize the variational energy (\ref{var_energy}). However, the ansatz (\ref{var_psi}) is inappropriate for immiscible and imbalanced droplets and supersolid states (see (\cite{bisset_2021_quantum}) for alternative ansatz).

\section{Density contrast}\label{C}
The groundstate phase diagrams for a balanced binary mixture are depicted in Fig. \ref{fig:1}. The binary mixture can be in one of the three phases: a SF state, a SS state with periodic density modulation, and a 2D array of isolated droplets. These distinct phases are best characterized by the density contrast, defined as \cite{bland_2022_alternatingdomain}
\begin{equation}
	\mathcal{C}=\frac{(n_{\rm max}-n_{\rm min})}{(n_{\rm max}+n_{\rm min})}.\label{contrast}
\end{equation}
 Here $n_{max}$ and $n_{min}$ are the neighboring maxima and minima as one moves on the x-y plane. A SF state corresponds to a smooth density distribution with $n_{\rm max}= n_{\rm min}$ which implies $\mathcal{C}=0$. In an insulating droplet state when there is no overlap between the droplets ($n_{\rm min}\approx 0$), the Eq. \ref{contrast} gives $\mathcal{C}\approx 1$. Whereas in a SS state, the droplets are connected by a low-density superfluid ($n_{\rm min}\neq 0$) and the density contrast $\mathcal{C}$ attains an intermediate value between 0 and 1. In this work, we consider \cite{bland_2022_twodimensional} $\mathcal{C}=0$ to be a superfluid phase, $0<\mathcal{C}<0.96$ to be a supersolid, and $\mathcal{C}>0.96$ to be a droplet state. In Fig. \ref{fig:contrast_phase_diagram}, we plot the different contrast ($\mathcal{C}$) regimes as a function of intra-, inter-species scattering length and the number of particles. 

\section{Effective potential}\label{D}
Each condensate experiences an effective potential due to the presence of the other component. The effective potential experienced by species-$i$ due to the presence of species-$j$  is given by,
\begin{align}
	V_{eff}^{ij}(\textbf{r})=&g_{ij}\abs{\psi_j(\textbf{r})}^2+ \int d\textbf{r}' V_{ij}^{dd}(\textbf{r}-\textbf{r}')\abs{\psi_j(\bf{r}^{\prime})}^2.
\end{align}
{\unskip\parfillskip 0pt\par}
\onecolumngrid

\begin{figure}[tb!]
	\centering
	\includegraphics[width=\textwidth]{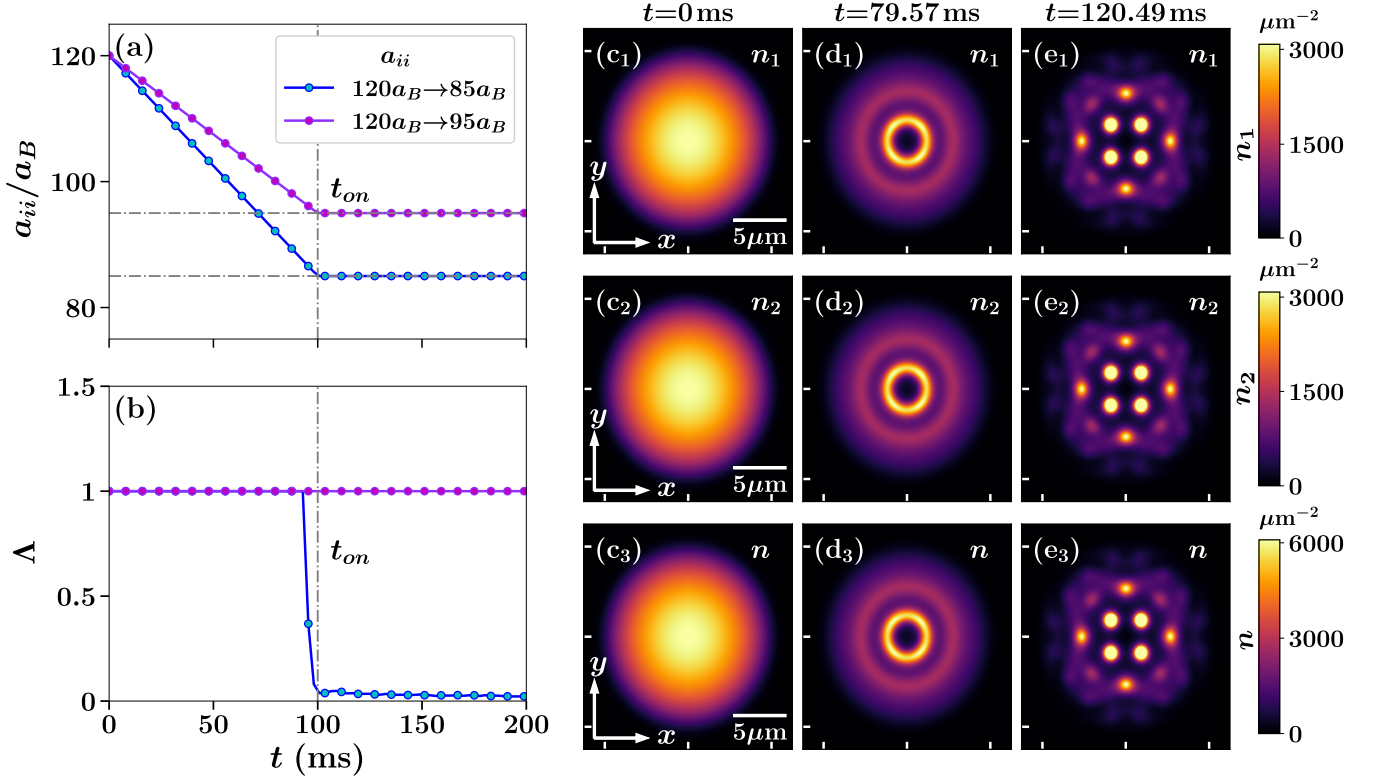}
	\caption{Quenching intra-species scattering lengths for a fixed values of $a_{12}=90a_B$. Panel (a) shows the linear ramp with ramp time $t_{on}=100$ms. From an initial value of $a_{ii}=120a_B$, the intra-species scattering lengths are quenched to a final value $a_{ii}=85a_B$ and $a_{ii}=95a_B$, respectively through this ramp. Panel (b) shows the corresponding variation of the overlap integral with time. The snapshots of the density profile of species-1 $n_1(x,y,z=0)$ ($c_1-e_1$), species-2 $n_2(x,y,z=0)$ ($c_2-e_2$), and of the total system $n(x,y,z=0)$ ($c_3-e_3$), following a quench from a miscible SF state ($a_{ii}=120a_B$) to a miscible SS state ($a_{ii}=95a_B$). Both the condensate consists of $N=6\times 10^4$ number atoms, and the balanced binary mixture is confined in a harmonic trap with $(\omega_x,\omega_y,\omega_z)=2\pi\times(45,45,133)$Hz. The colorbar denotes the density in units of $\mu \rm m^{-2}$.}\label{fig:Dynamics_quench_aii}
\end{figure}

\twocolumngrid
\noindent In the main text, we have shown the density profiles of an axially immiscible SD state in an imbalanced binary mixture (see Fig. \ref{fig:1D_chpt_den_imbalanced_sd}($\rm a_2$)). Where we observe that the major\footnote{In an imbalanced binary mixture, the species with a larger number of atoms is referred to as the major component and the other species as the minor component.} component with smaller $a_{ii}$ and a larger population acquires the central position and the minor component with comparatively larger $a_{ii}$ and a smaller number of atoms is bound at each end along the axial direction. Here in Fig. \ref{fig:2D_veff_immis_sd}, we have shown the corresponding effective potentials experienced by each species due to the presence of other species. The species-1 encountered a minimum potential at the trap center (which is elongated along the axial direction (z-axis)), whereas the second species experienced a maximum effective potential there but a minimum effective potential at each end of the minimum effective potential domain formed by the condensate-1.

\section{Quench dynamics}\label{E}
\begin{figure}[tb!]
	\centering
	\includegraphics[width=0.48\textwidth]{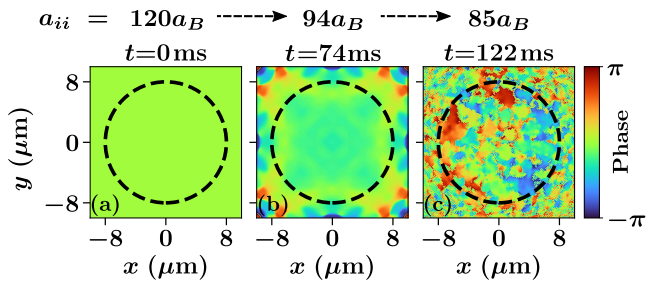}
	\caption{Characteristic phase profiles (a){-}(c) of the composite binary mixture in the x-y plane at a specific time instant following a linear quench with ramp time $100$ms from a miscible SF state with $a_{ii}=120a_B$ to an immiscible MD state with $a_{ii}=85a_B$ keeping the inter-species scattering length ($a_{12}$) fixed at $90a_B$. The black dashed circles delineate the edges of the dipolar binary mixture. The phase profiles correspond to (a) miscible SF state, (b) miscible SS state, and (c) immiscible MD state. The parameters are same as in the Fig. \ref{fig:Dyna_FFT_Quench_aii_120_85_t}.}\label{fig:phase_dyna_quench_aij_90_aii_120_85}
\end{figure}

To track the emergent features of the intra-species interaction quench as we discussed in the main text, here we have shown the time evolution of the density profiles and the phase of the binary mixture. In Figs. \ref{fig:Dynamics_quench_aii}(a), we have shown two different interaction quenches of $a_{ii}$, one from $120a_B$ to $95a_B$, and the other from $120a_B$ to $85a_B$. Following these interaction quenches, the corresponding time evolution of the overlap integral $\Lambda$ is shown in Fig. \ref{fig:Dynamics_quench_aii}(b). Initially, at $t=0$ ms the mixture forms a miscible SF state with a smooth 2D TF distribution (see Fig. \ref{fig:Dynamics_quench_aii}($\rm c_1-c_3$)) which corresponds to a global phase coherence as can be seen from Fig. \ref{fig:phase_dyna_quench_aij_90_aii_120_85}(a). However, since the quench is performed across the phase boundary, it excites the roton instability in the binary system leading to ring-shaped density structures as can be seen in Fig. \ref{fig:Dynamics_quench_aii}($\rm d_1-d_3$). The appearance of the roton mode is readily visible in the momentum space. Due to the circular symmetry of the trap geometry ($\omega_x=\omega_y=\omega$), the roton population is spread over a ring in the $k_x-k_y$ plane and for $k_y=0$ it corresponds to the appearance of two prominent side peaks as discussed in the main text (see Fig. \ref{fig:Dyna_FFT_Quench_aii_120_85}(b)).\par
As we decrease the intra-species scattering lengths more, the ring-shaped density structure breaks into several overlapping density humps (droplets) and the binary mixture forms a miscible SS state (Fig. \ref{fig:Dynamics_quench_aii}($\rm e_1-e_3$)). This SS state corresponds to almost a perfect global phase coherence with a very small fluctuation in the phase observed due to the interaction quench performed across the phase boundary (see Fig. \ref{fig:phase_dyna_quench_aij_90_aii_120_85}(b)). Instead of interaction quench by the evaporative cooling mechanism directly into the SS state, one could produce a SS state with robust global phase coherence as demonstrated in \cite{bland_2022_twodimensional}. Further decreasing the $a_{ii}$, the phase coherence between these droplets is completely lost (see Fig. \ref{fig:phase_dyna_quench_aij_90_aii_120_85}(c)) and the binary mixture forms a 2D array of immiscible MD crystal (not shown here).

\bibliographystyle{apsrev4-2}
\bibliography{reference} 

\providecommand{\noopsort}[1]{}
\begin{thebibliography}{84}%
\makeatletter
\providecommand \@ifxundefined [1]{%
 \@ifx{#1\undefined}
}%
\providecommand \@ifnum [1]{%
 \ifnum #1\expandafter \@firstoftwo
 \else \expandafter \@secondoftwo
 \fi
}%
\providecommand \@ifx [1]{%
 \ifx #1\expandafter \@firstoftwo
 \else \expandafter \@secondoftwo
 \fi
}%
\providecommand \natexlab [1]{#1}%
\providecommand \enquote  [1]{``#1''}%
\providecommand \bibnamefont  [1]{#1}%
\providecommand \bibfnamefont [1]{#1}%
\providecommand \citenamefont [1]{#1}%
\providecommand \href@noop [0]{\@secondoftwo}%
\providecommand \href [0]{\begingroup \@sanitize@url \@href}%
\providecommand \@href[1]{\@@startlink{#1}\@@href}%
\providecommand \@@href[1]{\endgroup#1\@@endlink}%
\providecommand \@sanitize@url [0]{\catcode `\\12\catcode `\$12\catcode
  `\&12\catcode `\#12\catcode `\^12\catcode `\_12\catcode `\%12\relax}%
\providecommand \@@startlink[1]{}%
\providecommand \@@endlink[0]{}%
\providecommand \url  [0]{\begingroup\@sanitize@url \@url }%
\providecommand \@url [1]{\endgroup\@href {#1}{\urlprefix }}%
\providecommand \urlprefix  [0]{URL }%
\providecommand \Eprint [0]{\href }%
\providecommand \doibase [0]{https://doi.org/}%
\providecommand \selectlanguage [0]{\@gobble}%
\providecommand \bibinfo  [0]{\@secondoftwo}%
\providecommand \bibfield  [0]{\@secondoftwo}%
\providecommand \translation [1]{[#1]}%
\providecommand \BibitemOpen [0]{}%
\providecommand \bibitemStop [0]{}%
\providecommand \bibitemNoStop [0]{.\EOS\space}%
\providecommand \EOS [0]{\spacefactor3000\relax}%
\providecommand \BibitemShut  [1]{\csname bibitem#1\endcsname}%
\let\auto@bib@innerbib\@empty
\bibitem [{\citenamefont {Petrov}(2015)}]{petrov_2015_quantum}%
  \BibitemOpen
  \bibfield  {author} {\bibinfo {author} {\bibfnamefont {D.~S.}\ \bibnamefont
  {Petrov}},\ }\href {https://doi.org/10.1103/PhysRevLett.115.155302}
  {\bibfield  {journal} {\bibinfo  {journal} {Physical Review Letters}\
  }\textbf {\bibinfo {volume} {115}},\ \bibinfo {pages} {155302} (\bibinfo
  {year} {2015})}\BibitemShut {NoStop}%
\bibitem [{\citenamefont {Lee}\ \emph {et~al.}(1957)\citenamefont {Lee},
  \citenamefont {Huang},\ and\ \citenamefont {Yang}}]{lee_1957_eigenvalues}%
  \BibitemOpen
  \bibfield  {author} {\bibinfo {author} {\bibfnamefont {T.~D.}\ \bibnamefont
  {Lee}}, \bibinfo {author} {\bibfnamefont {K.}~\bibnamefont {Huang}},\ and\
  \bibinfo {author} {\bibfnamefont {C.~N.}\ \bibnamefont {Yang}},\ }\href
  {https://doi.org/10.1103/PhysRev.106.1135} {\bibfield  {journal} {\bibinfo
  {journal} {Physical Review}\ }\textbf {\bibinfo {volume} {106}},\ \bibinfo
  {pages} {1135} (\bibinfo {year} {1957})}\BibitemShut {NoStop}%
\bibitem [{\citenamefont {Ilzh{\"o}fer}\ \emph {et~al.}(2021)\citenamefont
  {Ilzh{\"o}fer}, \citenamefont {Sohmen}, \citenamefont {Durastante},
  \citenamefont {Politi}, \citenamefont {Trautmann}, \citenamefont {Natale},
  \citenamefont {Morpurgo}, \citenamefont {Giamarchi}, \citenamefont {Chomaz},
  \citenamefont {Mark},\ and\ \citenamefont {Ferlaino}}]{ilzhofer_2021_phase}%
  \BibitemOpen
  \bibfield  {author} {\bibinfo {author} {\bibfnamefont {P.}~\bibnamefont
  {Ilzh{\"o}fer}}, \bibinfo {author} {\bibfnamefont {M.}~\bibnamefont
  {Sohmen}}, \bibinfo {author} {\bibfnamefont {G.}~\bibnamefont {Durastante}},
  \bibinfo {author} {\bibfnamefont {C.}~\bibnamefont {Politi}}, \bibinfo
  {author} {\bibfnamefont {A.}~\bibnamefont {Trautmann}}, \bibinfo {author}
  {\bibfnamefont {G.}~\bibnamefont {Natale}}, \bibinfo {author} {\bibfnamefont
  {G.}~\bibnamefont {Morpurgo}}, \bibinfo {author} {\bibfnamefont
  {T.}~\bibnamefont {Giamarchi}}, \bibinfo {author} {\bibfnamefont
  {L.}~\bibnamefont {Chomaz}}, \bibinfo {author} {\bibfnamefont {M.~J.}\
  \bibnamefont {Mark}},\ and\ \bibinfo {author} {\bibfnamefont
  {F.}~\bibnamefont {Ferlaino}},\ }\href
  {https://doi.org/10.1038/s41567-020-01100-3} {\bibfield  {journal} {\bibinfo
  {journal} {Nature Physics}\ }\textbf {\bibinfo {volume} {17}},\ \bibinfo
  {pages} {356} (\bibinfo {year} {2021})}\BibitemShut {NoStop}%
\bibitem [{\citenamefont {Tanzi}\ \emph
  {et~al.}(2019{\natexlab{a}})\citenamefont {Tanzi}, \citenamefont {Lucioni},
  \citenamefont {Fam{\`a}}, \citenamefont {Catani}, \citenamefont {Fioretti},
  \citenamefont {Gabbanini}, \citenamefont {Bisset}, \citenamefont {Santos},\
  and\ \citenamefont {Modugno}}]{tanzi_2019_observation}%
  \BibitemOpen
  \bibfield  {author} {\bibinfo {author} {\bibfnamefont {L.}~\bibnamefont
  {Tanzi}}, \bibinfo {author} {\bibfnamefont {E.}~\bibnamefont {Lucioni}},
  \bibinfo {author} {\bibfnamefont {F.}~\bibnamefont {Fam{\`a}}}, \bibinfo
  {author} {\bibfnamefont {J.}~\bibnamefont {Catani}}, \bibinfo {author}
  {\bibfnamefont {A.}~\bibnamefont {Fioretti}}, \bibinfo {author}
  {\bibfnamefont {C.}~\bibnamefont {Gabbanini}}, \bibinfo {author}
  {\bibfnamefont {R.~N.}\ \bibnamefont {Bisset}}, \bibinfo {author}
  {\bibfnamefont {L.}~\bibnamefont {Santos}},\ and\ \bibinfo {author}
  {\bibfnamefont {G.}~\bibnamefont {Modugno}},\ }\href
  {https://doi.org/10.1103/PhysRevLett.122.130405} {\bibfield  {journal}
  {\bibinfo  {journal} {Physical Review Letters}\ }\textbf {\bibinfo {volume}
  {122}},\ \bibinfo {pages} {130405} (\bibinfo {year}
  {2019}{\natexlab{a}})}\BibitemShut {NoStop}%
\bibitem [{\citenamefont {Toennies}\ \emph {et~al.}(2001)\citenamefont
  {Toennies}, \citenamefont {Vilesov},\ and\ \citenamefont
  {Whaley}}]{toennies_2001_superfluid}%
  \BibitemOpen
  \bibfield  {author} {\bibinfo {author} {\bibfnamefont {J.~P.}\ \bibnamefont
  {Toennies}}, \bibinfo {author} {\bibfnamefont {A.~F.}\ \bibnamefont
  {Vilesov}},\ and\ \bibinfo {author} {\bibfnamefont {K.~B.}\ \bibnamefont
  {Whaley}},\ }\href {https://doi.org/10.1063/1.1359707} {\bibfield  {journal}
  {\bibinfo  {journal} {Physics Today}\ }\textbf {\bibinfo {volume} {54}},\
  \bibinfo {pages} {31} (\bibinfo {year} {2001})}\BibitemShut {NoStop}%
\bibitem [{\citenamefont {Toennies}\ and\ \citenamefont
  {Vilesov}(2004)}]{toennies_2004_superfluid}%
  \BibitemOpen
  \bibfield  {author} {\bibinfo {author} {\bibfnamefont {J.~P.}\ \bibnamefont
  {Toennies}}\ and\ \bibinfo {author} {\bibfnamefont {A.~F.}\ \bibnamefont
  {Vilesov}},\ }\href {https://doi.org/10.1002/anie.200300611} {\bibfield
  {journal} {\bibinfo  {journal} {Angewandte Chemie International Edition}\
  }\textbf {\bibinfo {volume} {43}},\ \bibinfo {pages} {2622} (\bibinfo {year}
  {2004})}\BibitemShut {NoStop}%
\bibitem [{\citenamefont {Barranco}\ \emph {et~al.}(2006)\citenamefont
  {Barranco}, \citenamefont {Guardiola}, \citenamefont {Hern{\'a}ndez},
  \citenamefont {Mayol}, \citenamefont {Navarro},\ and\ \citenamefont
  {Pi}}]{barranco_2006_helium}%
  \BibitemOpen
  \bibfield  {author} {\bibinfo {author} {\bibfnamefont {M.}~\bibnamefont
  {Barranco}}, \bibinfo {author} {\bibfnamefont {R.}~\bibnamefont {Guardiola}},
  \bibinfo {author} {\bibfnamefont {S.}~\bibnamefont {Hern{\'a}ndez}}, \bibinfo
  {author} {\bibfnamefont {R.}~\bibnamefont {Mayol}}, \bibinfo {author}
  {\bibfnamefont {J.}~\bibnamefont {Navarro}},\ and\ \bibinfo {author}
  {\bibfnamefont {M.}~\bibnamefont {Pi}},\ }\href
  {https://doi.org/10.1007/s10909-005-9267-0} {\bibfield  {journal} {\bibinfo
  {journal} {Journal of Low Temperature Physics}\ }\textbf {\bibinfo {volume}
  {142}},\ \bibinfo {pages} {1} (\bibinfo {year} {2006})}\BibitemShut {NoStop}%
\bibitem [{\citenamefont {Ancilotto}\ \emph {et~al.}(2017)\citenamefont
  {Ancilotto}, \citenamefont {Barranco}, \citenamefont {Coppens}, \citenamefont
  {Eloranta}, \citenamefont {Halberstadt}, \citenamefont {Hernando},
  \citenamefont {Mateo},\ and\ \citenamefont {Pi}}]{ancilotto_2017_density}%
  \BibitemOpen
  \bibfield  {author} {\bibinfo {author} {\bibfnamefont {F.}~\bibnamefont
  {Ancilotto}}, \bibinfo {author} {\bibfnamefont {M.}~\bibnamefont {Barranco}},
  \bibinfo {author} {\bibfnamefont {F.}~\bibnamefont {Coppens}}, \bibinfo
  {author} {\bibfnamefont {J.}~\bibnamefont {Eloranta}}, \bibinfo {author}
  {\bibfnamefont {N.}~\bibnamefont {Halberstadt}}, \bibinfo {author}
  {\bibfnamefont {A.}~\bibnamefont {Hernando}}, \bibinfo {author}
  {\bibfnamefont {D.}~\bibnamefont {Mateo}},\ and\ \bibinfo {author}
  {\bibfnamefont {M.}~\bibnamefont {Pi}},\ }\href
  {https://doi.org/10.1080/0144235X.2017.1351672} {\bibfield  {journal}
  {\bibinfo  {journal} {International Reviews in Physical Chemistry}\ }\textbf
  {\bibinfo {volume} {36}},\ \bibinfo {pages} {621} (\bibinfo {year}
  {2017})}\BibitemShut {NoStop}%
\bibitem [{\citenamefont {Chin}\ \emph {et~al.}(2010)\citenamefont {Chin},
  \citenamefont {Grimm}, \citenamefont {Julienne},\ and\ \citenamefont
  {Tiesinga}}]{chin_2010_Feshbach}%
  \BibitemOpen
  \bibfield  {author} {\bibinfo {author} {\bibfnamefont {C.}~\bibnamefont
  {Chin}}, \bibinfo {author} {\bibfnamefont {R.}~\bibnamefont {Grimm}},
  \bibinfo {author} {\bibfnamefont {P.}~\bibnamefont {Julienne}},\ and\
  \bibinfo {author} {\bibfnamefont {E.}~\bibnamefont {Tiesinga}},\ }\href
  {https://doi.org/10.1103/RevModPhys.82.1225} {\bibfield  {journal} {\bibinfo
  {journal} {Rev. Mod. Phys.}\ }\textbf {\bibinfo {volume} {82}},\ \bibinfo
  {pages} {1225} (\bibinfo {year} {2010})}\BibitemShut {NoStop}%
\bibitem [{\citenamefont {Kadau}\ \emph {et~al.}(2016)\citenamefont {Kadau},
  \citenamefont {Schmitt}, \citenamefont {Wenzel}, \citenamefont {Wink},
  \citenamefont {Maier}, \citenamefont {{Ferrier-Barbut}},\ and\ \citenamefont
  {Pfau}}]{kadau_2016_observing}%
  \BibitemOpen
  \bibfield  {author} {\bibinfo {author} {\bibfnamefont {H.}~\bibnamefont
  {Kadau}}, \bibinfo {author} {\bibfnamefont {M.}~\bibnamefont {Schmitt}},
  \bibinfo {author} {\bibfnamefont {M.}~\bibnamefont {Wenzel}}, \bibinfo
  {author} {\bibfnamefont {C.}~\bibnamefont {Wink}}, \bibinfo {author}
  {\bibfnamefont {T.}~\bibnamefont {Maier}}, \bibinfo {author} {\bibfnamefont
  {I.}~\bibnamefont {{Ferrier-Barbut}}},\ and\ \bibinfo {author} {\bibfnamefont
  {T.}~\bibnamefont {Pfau}},\ }\href {https://doi.org/10.1038/nature16485}
  {\bibfield  {journal} {\bibinfo  {journal} {Nature}\ }\textbf {\bibinfo
  {volume} {530}},\ \bibinfo {pages} {194} (\bibinfo {year}
  {2016})}\BibitemShut {NoStop}%
\bibitem [{\citenamefont {W{\"a}chtler}\ and\ \citenamefont
  {Santos}(2016{\natexlab{a}})}]{wachtler_2016_groundstate}%
  \BibitemOpen
  \bibfield  {author} {\bibinfo {author} {\bibfnamefont {F.}~\bibnamefont
  {W{\"a}chtler}}\ and\ \bibinfo {author} {\bibfnamefont {L.}~\bibnamefont
  {Santos}},\ }\href {https://doi.org/10.1103/PhysRevA.94.043618} {\bibfield
  {journal} {\bibinfo  {journal} {Physical Review A}\ }\textbf {\bibinfo
  {volume} {94}},\ \bibinfo {pages} {043618} (\bibinfo {year}
  {2016}{\natexlab{a}})}\BibitemShut {NoStop}%
\bibitem [{\citenamefont {{Ferrier-Barbut}}\ \emph {et~al.}(2016)\citenamefont
  {{Ferrier-Barbut}}, \citenamefont {Kadau}, \citenamefont {Schmitt},
  \citenamefont {Wenzel},\ and\ \citenamefont
  {Pfau}}]{ferrier-barbut_2016_observation}%
  \BibitemOpen
  \bibfield  {author} {\bibinfo {author} {\bibfnamefont {I.}~\bibnamefont
  {{Ferrier-Barbut}}}, \bibinfo {author} {\bibfnamefont {H.}~\bibnamefont
  {Kadau}}, \bibinfo {author} {\bibfnamefont {M.}~\bibnamefont {Schmitt}},
  \bibinfo {author} {\bibfnamefont {M.}~\bibnamefont {Wenzel}},\ and\ \bibinfo
  {author} {\bibfnamefont {T.}~\bibnamefont {Pfau}},\ }\href
  {https://doi.org/10.1103/PhysRevLett.116.215301} {\bibfield  {journal}
  {\bibinfo  {journal} {Physical Review Letters}\ }\textbf {\bibinfo {volume}
  {116}},\ \bibinfo {pages} {215301} (\bibinfo {year} {2016})}\BibitemShut
  {NoStop}%
\bibitem [{\citenamefont
  {Chomaz}(2016)}]{chomaz_2016_quantumfluctuationdriven}%
  \BibitemOpen
  \bibfield  {author} {\bibinfo {author} {\bibfnamefont {L.}~\bibnamefont
  {Chomaz}},\ }\bibfield  {journal} {\bibinfo  {journal} {Physical Review X}\
  }\textbf {\bibinfo {volume} {6}},\ \href
  {https://doi.org/10.1103/PhysRevX.6.041039} {10.1103/PhysRevX.6.041039}
  (\bibinfo {year} {2016})\BibitemShut {NoStop}%
\bibitem [{\citenamefont {Petter}\ \emph {et~al.}(2019)\citenamefont {Petter},
  \citenamefont {Natale}, \citenamefont {{\noopsort{bijnen}}{van Bijnen}},
  \citenamefont {Patscheider}, \citenamefont {Mark}, \citenamefont {Chomaz},\
  and\ \citenamefont {Ferlaino}}]{petter_2019_probing}%
  \BibitemOpen
  \bibfield  {author} {\bibinfo {author} {\bibfnamefont {D.}~\bibnamefont
  {Petter}}, \bibinfo {author} {\bibfnamefont {G.}~\bibnamefont {Natale}},
  \bibinfo {author} {\bibfnamefont {R.~M.~W.}\ \bibnamefont
  {{\noopsort{bijnen}}{van Bijnen}}}, \bibinfo {author} {\bibfnamefont
  {A.}~\bibnamefont {Patscheider}}, \bibinfo {author} {\bibfnamefont {M.~J.}\
  \bibnamefont {Mark}}, \bibinfo {author} {\bibfnamefont {L.}~\bibnamefont
  {Chomaz}},\ and\ \bibinfo {author} {\bibfnamefont {F.}~\bibnamefont
  {Ferlaino}},\ }\href {https://doi.org/10.1103/PhysRevLett.122.183401}
  {\bibfield  {journal} {\bibinfo  {journal} {Physical Review Letters}\
  }\textbf {\bibinfo {volume} {122}},\ \bibinfo {pages} {183401} (\bibinfo
  {year} {2019})}\BibitemShut {NoStop}%
\bibitem [{\citenamefont {Chomaz}\ \emph {et~al.}(2019)\citenamefont {Chomaz},
  \citenamefont {Petter}, \citenamefont {Ilzh{\"o}fer}, \citenamefont {Natale},
  \citenamefont {Trautmann}, \citenamefont {Politi}, \citenamefont
  {Durastante}, \citenamefont {{\noopsort{bijnen}}{van Bijnen}}, \citenamefont
  {Patscheider}, \citenamefont {Sohmen}, \citenamefont {Mark},\ and\
  \citenamefont {Ferlaino}}]{chomaz_2019_longlived}%
  \BibitemOpen
  \bibfield  {author} {\bibinfo {author} {\bibfnamefont {L.}~\bibnamefont
  {Chomaz}}, \bibinfo {author} {\bibfnamefont {D.}~\bibnamefont {Petter}},
  \bibinfo {author} {\bibfnamefont {P.}~\bibnamefont {Ilzh{\"o}fer}}, \bibinfo
  {author} {\bibfnamefont {G.}~\bibnamefont {Natale}}, \bibinfo {author}
  {\bibfnamefont {A.}~\bibnamefont {Trautmann}}, \bibinfo {author}
  {\bibfnamefont {C.}~\bibnamefont {Politi}}, \bibinfo {author} {\bibfnamefont
  {G.}~\bibnamefont {Durastante}}, \bibinfo {author} {\bibfnamefont {R.~M.~W.}\
  \bibnamefont {{\noopsort{bijnen}}{van Bijnen}}}, \bibinfo {author}
  {\bibfnamefont {A.}~\bibnamefont {Patscheider}}, \bibinfo {author}
  {\bibfnamefont {M.}~\bibnamefont {Sohmen}}, \bibinfo {author} {\bibfnamefont
  {M.~J.}\ \bibnamefont {Mark}},\ and\ \bibinfo {author} {\bibfnamefont
  {F.}~\bibnamefont {Ferlaino}},\ }\href
  {https://doi.org/10.1103/PhysRevX.9.021012} {\bibfield  {journal} {\bibinfo
  {journal} {Physical Review X}\ }\textbf {\bibinfo {volume} {9}},\ \bibinfo
  {pages} {021012} (\bibinfo {year} {2019})}\BibitemShut {NoStop}%
\bibitem [{\citenamefont {Schmitt}\ \emph {et~al.}(2016)\citenamefont
  {Schmitt}, \citenamefont {Wenzel}, \citenamefont {B{\"o}ttcher},
  \citenamefont {{Ferrier-Barbut}},\ and\ \citenamefont
  {Pfau}}]{schmitt_2016_selfbound}%
  \BibitemOpen
  \bibfield  {author} {\bibinfo {author} {\bibfnamefont {M.}~\bibnamefont
  {Schmitt}}, \bibinfo {author} {\bibfnamefont {M.}~\bibnamefont {Wenzel}},
  \bibinfo {author} {\bibfnamefont {F.}~\bibnamefont {B{\"o}ttcher}}, \bibinfo
  {author} {\bibfnamefont {I.}~\bibnamefont {{Ferrier-Barbut}}},\ and\ \bibinfo
  {author} {\bibfnamefont {T.}~\bibnamefont {Pfau}},\ }\href
  {https://doi.org/10.1038/nature20126} {\bibfield  {journal} {\bibinfo
  {journal} {Nature}\ }\textbf {\bibinfo {volume} {539}},\ \bibinfo {pages}
  {259} (\bibinfo {year} {2016})}\BibitemShut {NoStop}%
\bibitem [{\citenamefont {Bisset}\ \emph {et~al.}(2016)\citenamefont {Bisset},
  \citenamefont {Wilson}, \citenamefont {Baillie},\ and\ \citenamefont
  {Blakie}}]{bisset_2016_groundstate}%
  \BibitemOpen
  \bibfield  {author} {\bibinfo {author} {\bibfnamefont {R.~N.}\ \bibnamefont
  {Bisset}}, \bibinfo {author} {\bibfnamefont {R.~M.}\ \bibnamefont {Wilson}},
  \bibinfo {author} {\bibfnamefont {D.}~\bibnamefont {Baillie}},\ and\ \bibinfo
  {author} {\bibfnamefont {P.~B.}\ \bibnamefont {Blakie}},\ }\href
  {https://doi.org/10.1103/PhysRevA.94.033619} {\bibfield  {journal} {\bibinfo
  {journal} {Physical Review A}\ }\textbf {\bibinfo {volume} {94}},\ \bibinfo
  {pages} {033619} (\bibinfo {year} {2016})}\BibitemShut {NoStop}%
\bibitem [{\citenamefont {Baillie}\ \emph {et~al.}(2017)\citenamefont
  {Baillie}, \citenamefont {Wilson},\ and\ \citenamefont
  {Blakie}}]{baillie_2017_collective}%
  \BibitemOpen
  \bibfield  {author} {\bibinfo {author} {\bibfnamefont {D.}~\bibnamefont
  {Baillie}}, \bibinfo {author} {\bibfnamefont {R.~M.}\ \bibnamefont
  {Wilson}},\ and\ \bibinfo {author} {\bibfnamefont {P.~B.}\ \bibnamefont
  {Blakie}},\ }\href {https://doi.org/10.1103/PhysRevLett.119.255302}
  {\bibfield  {journal} {\bibinfo  {journal} {Physical Review Letters}\
  }\textbf {\bibinfo {volume} {119}},\ \bibinfo {pages} {255302} (\bibinfo
  {year} {2017})}\BibitemShut {NoStop}%
\bibitem [{\citenamefont {Cinti}\ and\ \citenamefont
  {Boninsegni}(2017)}]{cinti_2017_classical}%
  \BibitemOpen
  \bibfield  {author} {\bibinfo {author} {\bibfnamefont {F.}~\bibnamefont
  {Cinti}}\ and\ \bibinfo {author} {\bibfnamefont {M.}~\bibnamefont
  {Boninsegni}},\ }\href {https://doi.org/10.1103/PhysRevA.96.013627}
  {\bibfield  {journal} {\bibinfo  {journal} {Physical Review A}\ }\textbf
  {\bibinfo {volume} {96}},\ \bibinfo {pages} {013627} (\bibinfo {year}
  {2017})}\BibitemShut {NoStop}%
\bibitem [{\citenamefont {Baillie}\ and\ \citenamefont
  {Blakie}(2018)}]{baillie_2018_droplet}%
  \BibitemOpen
  \bibfield  {author} {\bibinfo {author} {\bibfnamefont {D.}~\bibnamefont
  {Baillie}}\ and\ \bibinfo {author} {\bibfnamefont {P.~B.}\ \bibnamefont
  {Blakie}},\ }\href {https://doi.org/10.1103/PhysRevLett.121.195301}
  {\bibfield  {journal} {\bibinfo  {journal} {Physical Review Letters}\
  }\textbf {\bibinfo {volume} {121}},\ \bibinfo {pages} {195301} (\bibinfo
  {year} {2018})}\BibitemShut {NoStop}%
\bibitem [{\citenamefont {Lee}\ \emph {et~al.}(2021{\natexlab{a}})\citenamefont
  {Lee}, \citenamefont {Baillie},\ and\ \citenamefont
  {Blakie}}]{lee_2021_numerical}%
  \BibitemOpen
  \bibfield  {author} {\bibinfo {author} {\bibfnamefont {A.-C.}\ \bibnamefont
  {Lee}}, \bibinfo {author} {\bibfnamefont {D.}~\bibnamefont {Baillie}},\ and\
  \bibinfo {author} {\bibfnamefont {P.~B.}\ \bibnamefont {Blakie}},\ }\href
  {https://doi.org/10.1103/PhysRevResearch.3.013283} {\bibfield  {journal}
  {\bibinfo  {journal} {Physical Review Research}\ }\textbf {\bibinfo {volume}
  {3}},\ \bibinfo {pages} {013283} (\bibinfo {year}
  {2021}{\natexlab{a}})}\BibitemShut {NoStop}%
\bibitem [{\citenamefont {W{\"a}chtler}\ and\ \citenamefont
  {Santos}(2016{\natexlab{b}})}]{wachtler_2016_quantum}%
  \BibitemOpen
  \bibfield  {author} {\bibinfo {author} {\bibfnamefont {F.}~\bibnamefont
  {W{\"a}chtler}}\ and\ \bibinfo {author} {\bibfnamefont {L.}~\bibnamefont
  {Santos}},\ }\href {https://doi.org/10.1103/PhysRevA.93.061603} {\bibfield
  {journal} {\bibinfo  {journal} {Physical Review A}\ }\textbf {\bibinfo
  {volume} {93}},\ \bibinfo {pages} {061603} (\bibinfo {year}
  {2016}{\natexlab{b}})}\BibitemShut {NoStop}%
\bibitem [{\citenamefont {Norcia}\ \emph {et~al.}(2021)\citenamefont {Norcia},
  \citenamefont {Politi}, \citenamefont {Klaus}, \citenamefont {Poli},
  \citenamefont {Sohmen}, \citenamefont {Mark}, \citenamefont {Bisset},
  \citenamefont {Santos},\ and\ \citenamefont
  {Ferlaino}}]{norcia_2021_twodimensional}%
  \BibitemOpen
  \bibfield  {author} {\bibinfo {author} {\bibfnamefont {M.~A.}\ \bibnamefont
  {Norcia}}, \bibinfo {author} {\bibfnamefont {C.}~\bibnamefont {Politi}},
  \bibinfo {author} {\bibfnamefont {L.}~\bibnamefont {Klaus}}, \bibinfo
  {author} {\bibfnamefont {E.}~\bibnamefont {Poli}}, \bibinfo {author}
  {\bibfnamefont {M.}~\bibnamefont {Sohmen}}, \bibinfo {author} {\bibfnamefont
  {M.~J.}\ \bibnamefont {Mark}}, \bibinfo {author} {\bibfnamefont {R.~N.}\
  \bibnamefont {Bisset}}, \bibinfo {author} {\bibfnamefont {L.}~\bibnamefont
  {Santos}},\ and\ \bibinfo {author} {\bibfnamefont {F.}~\bibnamefont
  {Ferlaino}},\ }\href {https://doi.org/10.1038/s41586-021-03725-7} {\bibfield
  {journal} {\bibinfo  {journal} {Nature}\ }\textbf {\bibinfo {volume} {596}},\
  \bibinfo {pages} {357} (\bibinfo {year} {2021})}\BibitemShut {NoStop}%
\bibitem [{\citenamefont {Bland}\ \emph
  {et~al.}(2022{\natexlab{a}})\citenamefont {Bland}, \citenamefont {Poli},
  \citenamefont {Politi}, \citenamefont {Klaus}, \citenamefont {Norcia},
  \citenamefont {Ferlaino}, \citenamefont {Santos},\ and\ \citenamefont
  {Bisset}}]{bland_2022_twodimensional}%
  \BibitemOpen
  \bibfield  {author} {\bibinfo {author} {\bibfnamefont {T.}~\bibnamefont
  {Bland}}, \bibinfo {author} {\bibfnamefont {E.}~\bibnamefont {Poli}},
  \bibinfo {author} {\bibfnamefont {C.}~\bibnamefont {Politi}}, \bibinfo
  {author} {\bibfnamefont {L.}~\bibnamefont {Klaus}}, \bibinfo {author}
  {\bibfnamefont {M.~A.}\ \bibnamefont {Norcia}}, \bibinfo {author}
  {\bibfnamefont {F.}~\bibnamefont {Ferlaino}}, \bibinfo {author}
  {\bibfnamefont {L.}~\bibnamefont {Santos}},\ and\ \bibinfo {author}
  {\bibfnamefont {R.~N.}\ \bibnamefont {Bisset}},\ }\href
  {https://doi.org/10.1103/PhysRevLett.128.195302} {\bibfield  {journal}
  {\bibinfo  {journal} {Physical Review Letters}\ }\textbf {\bibinfo {volume}
  {128}},\ \bibinfo {pages} {195302} (\bibinfo {year}
  {2022}{\natexlab{a}})}\BibitemShut {NoStop}%
\bibitem [{\citenamefont {Zhang}\ \emph {et~al.}(2019)\citenamefont {Zhang},
  \citenamefont {Maucher},\ and\ \citenamefont
  {Pohl}}]{zhang_2019_supersolidity}%
  \BibitemOpen
  \bibfield  {author} {\bibinfo {author} {\bibfnamefont {Y.-C.}\ \bibnamefont
  {Zhang}}, \bibinfo {author} {\bibfnamefont {F.}~\bibnamefont {Maucher}},\
  and\ \bibinfo {author} {\bibfnamefont {T.}~\bibnamefont {Pohl}},\ }\href
  {https://doi.org/10.1103/PhysRevLett.123.015301} {\bibfield  {journal}
  {\bibinfo  {journal} {Physical Review Letters}\ }\textbf {\bibinfo {volume}
  {123}},\ \bibinfo {pages} {015301} (\bibinfo {year} {2019})}\BibitemShut
  {NoStop}%
\bibitem [{\citenamefont {Roccuzzo}\ and\ \citenamefont
  {Ancilotto}(2019)}]{roccuzzo_2019_supersolid}%
  \BibitemOpen
  \bibfield  {author} {\bibinfo {author} {\bibfnamefont {S.~M.}\ \bibnamefont
  {Roccuzzo}}\ and\ \bibinfo {author} {\bibfnamefont {F.}~\bibnamefont
  {Ancilotto}},\ }\href {https://doi.org/10.1103/PhysRevA.99.041601} {\bibfield
   {journal} {\bibinfo  {journal} {Physical Review A}\ }\textbf {\bibinfo
  {volume} {99}},\ \bibinfo {pages} {041601} (\bibinfo {year}
  {2019})}\BibitemShut {NoStop}%
\bibitem [{\citenamefont {Poli}\ \emph {et~al.}(2021)\citenamefont {Poli},
  \citenamefont {Bland}, \citenamefont {Politi}, \citenamefont {Klaus},
  \citenamefont {Norcia}, \citenamefont {Ferlaino}, \citenamefont {Bisset},\
  and\ \citenamefont {Santos}}]{poli_2021_maintaining}%
  \BibitemOpen
  \bibfield  {author} {\bibinfo {author} {\bibfnamefont {E.}~\bibnamefont
  {Poli}}, \bibinfo {author} {\bibfnamefont {T.}~\bibnamefont {Bland}},
  \bibinfo {author} {\bibfnamefont {C.}~\bibnamefont {Politi}}, \bibinfo
  {author} {\bibfnamefont {L.}~\bibnamefont {Klaus}}, \bibinfo {author}
  {\bibfnamefont {M.~A.}\ \bibnamefont {Norcia}}, \bibinfo {author}
  {\bibfnamefont {F.}~\bibnamefont {Ferlaino}}, \bibinfo {author}
  {\bibfnamefont {R.~N.}\ \bibnamefont {Bisset}},\ and\ \bibinfo {author}
  {\bibfnamefont {L.}~\bibnamefont {Santos}},\ }\href
  {https://doi.org/10.1103/PhysRevA.104.063307} {\bibfield  {journal} {\bibinfo
   {journal} {Physical Review A}\ }\textbf {\bibinfo {volume} {104}},\ \bibinfo
  {pages} {063307} (\bibinfo {year} {2021})}\BibitemShut {NoStop}%
\bibitem [{\citenamefont {Natale}\ \emph {et~al.}(2019)\citenamefont {Natale},
  \citenamefont {{\noopsort{bijnen}}{van Bijnen}}, \citenamefont {Patscheider},
  \citenamefont {Petter}, \citenamefont {Mark}, \citenamefont {Chomaz},\ and\
  \citenamefont {Ferlaino}}]{natale_2019_excitation}%
  \BibitemOpen
  \bibfield  {author} {\bibinfo {author} {\bibfnamefont {G.}~\bibnamefont
  {Natale}}, \bibinfo {author} {\bibfnamefont {R.~M.~W.}\ \bibnamefont
  {{\noopsort{bijnen}}{van Bijnen}}}, \bibinfo {author} {\bibfnamefont
  {A.}~\bibnamefont {Patscheider}}, \bibinfo {author} {\bibfnamefont
  {D.}~\bibnamefont {Petter}}, \bibinfo {author} {\bibfnamefont {M.~J.}\
  \bibnamefont {Mark}}, \bibinfo {author} {\bibfnamefont {L.}~\bibnamefont
  {Chomaz}},\ and\ \bibinfo {author} {\bibfnamefont {F.}~\bibnamefont
  {Ferlaino}},\ }\href {https://doi.org/10.1103/PhysRevLett.123.050402}
  {\bibfield  {journal} {\bibinfo  {journal} {Physical Review Letters}\
  }\textbf {\bibinfo {volume} {123}},\ \bibinfo {pages} {050402} (\bibinfo
  {year} {2019})}\BibitemShut {NoStop}%
\bibitem [{\citenamefont {Hertkorn}\ \emph
  {et~al.}(2021{\natexlab{a}})\citenamefont {Hertkorn}, \citenamefont
  {Schmidt}, \citenamefont {Guo}, \citenamefont {B{\"o}ttcher}, \citenamefont
  {Ng}, \citenamefont {Graham}, \citenamefont {Uerlings}, \citenamefont
  {Langen}, \citenamefont {Zwierlein},\ and\ \citenamefont
  {Pfau}}]{hertkorn_2021_pattern}%
  \BibitemOpen
  \bibfield  {author} {\bibinfo {author} {\bibfnamefont {J.}~\bibnamefont
  {Hertkorn}}, \bibinfo {author} {\bibfnamefont {J.-N.}\ \bibnamefont
  {Schmidt}}, \bibinfo {author} {\bibfnamefont {M.}~\bibnamefont {Guo}},
  \bibinfo {author} {\bibfnamefont {F.}~\bibnamefont {B{\"o}ttcher}}, \bibinfo
  {author} {\bibfnamefont {K.~S.~H.}\ \bibnamefont {Ng}}, \bibinfo {author}
  {\bibfnamefont {S.~D.}\ \bibnamefont {Graham}}, \bibinfo {author}
  {\bibfnamefont {P.}~\bibnamefont {Uerlings}}, \bibinfo {author}
  {\bibfnamefont {T.}~\bibnamefont {Langen}}, \bibinfo {author} {\bibfnamefont
  {M.}~\bibnamefont {Zwierlein}},\ and\ \bibinfo {author} {\bibfnamefont
  {T.}~\bibnamefont {Pfau}},\ }\href
  {https://doi.org/10.1103/PhysRevResearch.3.033125} {\bibfield  {journal}
  {\bibinfo  {journal} {Physical Review Research}\ }\textbf {\bibinfo {volume}
  {3}},\ \bibinfo {pages} {033125} (\bibinfo {year}
  {2021}{\natexlab{a}})}\BibitemShut {NoStop}%
\bibitem [{\citenamefont {Sohmen}\ \emph {et~al.}(2021)\citenamefont {Sohmen},
  \citenamefont {Politi}, \citenamefont {Klaus}, \citenamefont {Chomaz},
  \citenamefont {Mark}, \citenamefont {Norcia},\ and\ \citenamefont
  {Ferlaino}}]{sohmen_2021_birth}%
  \BibitemOpen
  \bibfield  {author} {\bibinfo {author} {\bibfnamefont {M.}~\bibnamefont
  {Sohmen}}, \bibinfo {author} {\bibfnamefont {C.}~\bibnamefont {Politi}},
  \bibinfo {author} {\bibfnamefont {L.}~\bibnamefont {Klaus}}, \bibinfo
  {author} {\bibfnamefont {L.}~\bibnamefont {Chomaz}}, \bibinfo {author}
  {\bibfnamefont {M.~J.}\ \bibnamefont {Mark}}, \bibinfo {author}
  {\bibfnamefont {M.~A.}\ \bibnamefont {Norcia}},\ and\ \bibinfo {author}
  {\bibfnamefont {F.}~\bibnamefont {Ferlaino}},\ }\href
  {https://doi.org/10.1103/PhysRevLett.126.233401} {\bibfield  {journal}
  {\bibinfo  {journal} {Physical Review Letters}\ }\textbf {\bibinfo {volume}
  {126}},\ \bibinfo {pages} {233401} (\bibinfo {year} {2021})}\BibitemShut
  {NoStop}%
\bibitem [{\citenamefont {L{\'e}onard}\ \emph {et~al.}(2017)\citenamefont
  {L{\'e}onard}, \citenamefont {Morales}, \citenamefont {Zupancic},
  \citenamefont {Esslinger},\ and\ \citenamefont
  {Donner}}]{leonard_2017_supersolid}%
  \BibitemOpen
  \bibfield  {author} {\bibinfo {author} {\bibfnamefont {J.}~\bibnamefont
  {L{\'e}onard}}, \bibinfo {author} {\bibfnamefont {A.}~\bibnamefont
  {Morales}}, \bibinfo {author} {\bibfnamefont {P.}~\bibnamefont {Zupancic}},
  \bibinfo {author} {\bibfnamefont {T.}~\bibnamefont {Esslinger}},\ and\
  \bibinfo {author} {\bibfnamefont {T.}~\bibnamefont {Donner}},\ }\href
  {https://doi.org/10.1038/nature21067} {\bibfield  {journal} {\bibinfo
  {journal} {Nature}\ }\textbf {\bibinfo {volume} {543}},\ \bibinfo {pages}
  {87} (\bibinfo {year} {2017})}\BibitemShut {NoStop}%
\bibitem [{\citenamefont {Tanzi}\ \emph
  {et~al.}(2019{\natexlab{b}})\citenamefont {Tanzi}, \citenamefont {Roccuzzo},
  \citenamefont {Lucioni}, \citenamefont {Fam{\`a}}, \citenamefont {Fioretti},
  \citenamefont {Gabbanini}, \citenamefont {Modugno}, \citenamefont {Recati},\
  and\ \citenamefont {Stringari}}]{tanzi_2019_supersolid}%
  \BibitemOpen
  \bibfield  {author} {\bibinfo {author} {\bibfnamefont {L.}~\bibnamefont
  {Tanzi}}, \bibinfo {author} {\bibfnamefont {S.~M.}\ \bibnamefont {Roccuzzo}},
  \bibinfo {author} {\bibfnamefont {E.}~\bibnamefont {Lucioni}}, \bibinfo
  {author} {\bibfnamefont {F.}~\bibnamefont {Fam{\`a}}}, \bibinfo {author}
  {\bibfnamefont {A.}~\bibnamefont {Fioretti}}, \bibinfo {author}
  {\bibfnamefont {C.}~\bibnamefont {Gabbanini}}, \bibinfo {author}
  {\bibfnamefont {G.}~\bibnamefont {Modugno}}, \bibinfo {author} {\bibfnamefont
  {A.}~\bibnamefont {Recati}},\ and\ \bibinfo {author} {\bibfnamefont
  {S.}~\bibnamefont {Stringari}},\ }\href
  {https://doi.org/10.1038/s41586-019-1568-6} {\bibfield  {journal} {\bibinfo
  {journal} {Nature}\ }\textbf {\bibinfo {volume} {574}},\ \bibinfo {pages}
  {382} (\bibinfo {year} {2019}{\natexlab{b}})}\BibitemShut {NoStop}%
\bibitem [{\citenamefont {Guo}\ \emph {et~al.}(2019)\citenamefont {Guo},
  \citenamefont {B{\"o}ttcher}, \citenamefont {Hertkorn}, \citenamefont
  {Schmidt}, \citenamefont {Wenzel}, \citenamefont {B{\"u}chler}, \citenamefont
  {Langen},\ and\ \citenamefont {Pfau}}]{guo_2019_lowenergy}%
  \BibitemOpen
  \bibfield  {author} {\bibinfo {author} {\bibfnamefont {M.}~\bibnamefont
  {Guo}}, \bibinfo {author} {\bibfnamefont {F.}~\bibnamefont {B{\"o}ttcher}},
  \bibinfo {author} {\bibfnamefont {J.}~\bibnamefont {Hertkorn}}, \bibinfo
  {author} {\bibfnamefont {J.-N.}\ \bibnamefont {Schmidt}}, \bibinfo {author}
  {\bibfnamefont {M.}~\bibnamefont {Wenzel}}, \bibinfo {author} {\bibfnamefont
  {H.~P.}\ \bibnamefont {B{\"u}chler}}, \bibinfo {author} {\bibfnamefont
  {T.}~\bibnamefont {Langen}},\ and\ \bibinfo {author} {\bibfnamefont
  {T.}~\bibnamefont {Pfau}},\ }\href
  {https://doi.org/10.1038/s41586-019-1569-5} {\bibfield  {journal} {\bibinfo
  {journal} {Nature}\ }\textbf {\bibinfo {volume} {574}},\ \bibinfo {pages}
  {386} (\bibinfo {year} {2019})}\BibitemShut {NoStop}%
\bibitem [{\citenamefont {B{\"o}ttcher}\ \emph {et~al.}(2019)\citenamefont
  {B{\"o}ttcher}, \citenamefont {Schmidt}, \citenamefont {Wenzel},
  \citenamefont {Hertkorn}, \citenamefont {Guo}, \citenamefont {Langen},\ and\
  \citenamefont {Pfau}}]{bottcher_2019_transient}%
  \BibitemOpen
  \bibfield  {author} {\bibinfo {author} {\bibfnamefont {F.}~\bibnamefont
  {B{\"o}ttcher}}, \bibinfo {author} {\bibfnamefont {J.-N.}\ \bibnamefont
  {Schmidt}}, \bibinfo {author} {\bibfnamefont {M.}~\bibnamefont {Wenzel}},
  \bibinfo {author} {\bibfnamefont {J.}~\bibnamefont {Hertkorn}}, \bibinfo
  {author} {\bibfnamefont {M.}~\bibnamefont {Guo}}, \bibinfo {author}
  {\bibfnamefont {T.}~\bibnamefont {Langen}},\ and\ \bibinfo {author}
  {\bibfnamefont {T.}~\bibnamefont {Pfau}},\ }\href
  {https://doi.org/10.1103/PhysRevX.9.011051} {\bibfield  {journal} {\bibinfo
  {journal} {Physical Review X}\ }\textbf {\bibinfo {volume} {9}},\ \bibinfo
  {pages} {011051} (\bibinfo {year} {2019})}\BibitemShut {NoStop}%
\bibitem [{\citenamefont {Cabrera}\ \emph {et~al.}(2018)\citenamefont
  {Cabrera}, \citenamefont {Tanzi}, \citenamefont {Sanz}, \citenamefont
  {Naylor}, \citenamefont {Thomas}, \citenamefont {Cheiney},\ and\
  \citenamefont {Tarruell}}]{cabrera_2018_quantum}%
  \BibitemOpen
  \bibfield  {author} {\bibinfo {author} {\bibfnamefont {C.~R.}\ \bibnamefont
  {Cabrera}}, \bibinfo {author} {\bibfnamefont {L.}~\bibnamefont {Tanzi}},
  \bibinfo {author} {\bibfnamefont {J.}~\bibnamefont {Sanz}}, \bibinfo {author}
  {\bibfnamefont {B.}~\bibnamefont {Naylor}}, \bibinfo {author} {\bibfnamefont
  {P.}~\bibnamefont {Thomas}}, \bibinfo {author} {\bibfnamefont
  {P.}~\bibnamefont {Cheiney}},\ and\ \bibinfo {author} {\bibfnamefont
  {L.}~\bibnamefont {Tarruell}},\ }\href
  {https://doi.org/10.1126/science.aao5686} {\bibfield  {journal} {\bibinfo
  {journal} {Science}\ }\textbf {\bibinfo {volume} {359}},\ \bibinfo {pages}
  {301} (\bibinfo {year} {2018})}\BibitemShut {NoStop}%
\bibitem [{\citenamefont {Semeghini}\ \emph {et~al.}(2018)\citenamefont
  {Semeghini}, \citenamefont {Ferioli}, \citenamefont {Masi}, \citenamefont
  {Mazzinghi}, \citenamefont {Wolswijk}, \citenamefont {Minardi}, \citenamefont
  {Modugno}, \citenamefont {Modugno}, \citenamefont {Inguscio},\ and\
  \citenamefont {Fattori}}]{semeghini_2018_selfbound}%
  \BibitemOpen
  \bibfield  {author} {\bibinfo {author} {\bibfnamefont {G.}~\bibnamefont
  {Semeghini}}, \bibinfo {author} {\bibfnamefont {G.}~\bibnamefont {Ferioli}},
  \bibinfo {author} {\bibfnamefont {L.}~\bibnamefont {Masi}}, \bibinfo {author}
  {\bibfnamefont {C.}~\bibnamefont {Mazzinghi}}, \bibinfo {author}
  {\bibfnamefont {L.}~\bibnamefont {Wolswijk}}, \bibinfo {author}
  {\bibfnamefont {F.}~\bibnamefont {Minardi}}, \bibinfo {author} {\bibfnamefont
  {M.}~\bibnamefont {Modugno}}, \bibinfo {author} {\bibfnamefont
  {G.}~\bibnamefont {Modugno}}, \bibinfo {author} {\bibfnamefont
  {M.}~\bibnamefont {Inguscio}},\ and\ \bibinfo {author} {\bibfnamefont
  {M.}~\bibnamefont {Fattori}},\ }\href
  {https://doi.org/10.1103/PhysRevLett.120.235301} {\bibfield  {journal}
  {\bibinfo  {journal} {Physical Review Letters}\ }\textbf {\bibinfo {volume}
  {120}},\ \bibinfo {pages} {235301} (\bibinfo {year} {2018})}\BibitemShut
  {NoStop}%
\bibitem [{\citenamefont {Cheiney}\ \emph {et~al.}(2018)\citenamefont
  {Cheiney}, \citenamefont {Cabrera}, \citenamefont {Sanz}, \citenamefont
  {Naylor}, \citenamefont {Tanzi},\ and\ \citenamefont
  {Tarruell}}]{cheiney_2018_bright}%
  \BibitemOpen
  \bibfield  {author} {\bibinfo {author} {\bibfnamefont {P.}~\bibnamefont
  {Cheiney}}, \bibinfo {author} {\bibfnamefont {C.~R.}\ \bibnamefont
  {Cabrera}}, \bibinfo {author} {\bibfnamefont {J.}~\bibnamefont {Sanz}},
  \bibinfo {author} {\bibfnamefont {B.}~\bibnamefont {Naylor}}, \bibinfo
  {author} {\bibfnamefont {L.}~\bibnamefont {Tanzi}},\ and\ \bibinfo {author}
  {\bibfnamefont {L.}~\bibnamefont {Tarruell}},\ }\href
  {https://doi.org/10.1103/PhysRevLett.120.135301} {\bibfield  {journal}
  {\bibinfo  {journal} {Physical Review Letters}\ }\textbf {\bibinfo {volume}
  {120}},\ \bibinfo {pages} {135301} (\bibinfo {year} {2018})}\BibitemShut
  {NoStop}%
\bibitem [{\citenamefont {Ferioli}\ \emph {et~al.}(2019)\citenamefont
  {Ferioli}, \citenamefont {Semeghini}, \citenamefont {Masi}, \citenamefont
  {Giusti}, \citenamefont {Modugno}, \citenamefont {Inguscio}, \citenamefont
  {Gallem{\'i}}, \citenamefont {Recati},\ and\ \citenamefont
  {Fattori}}]{ferioli_2019_collisions}%
  \BibitemOpen
  \bibfield  {author} {\bibinfo {author} {\bibfnamefont {G.}~\bibnamefont
  {Ferioli}}, \bibinfo {author} {\bibfnamefont {G.}~\bibnamefont {Semeghini}},
  \bibinfo {author} {\bibfnamefont {L.}~\bibnamefont {Masi}}, \bibinfo {author}
  {\bibfnamefont {G.}~\bibnamefont {Giusti}}, \bibinfo {author} {\bibfnamefont
  {G.}~\bibnamefont {Modugno}}, \bibinfo {author} {\bibfnamefont
  {M.}~\bibnamefont {Inguscio}}, \bibinfo {author} {\bibfnamefont
  {A.}~\bibnamefont {Gallem{\'i}}}, \bibinfo {author} {\bibfnamefont
  {A.}~\bibnamefont {Recati}},\ and\ \bibinfo {author} {\bibfnamefont
  {M.}~\bibnamefont {Fattori}},\ }\href
  {https://doi.org/10.1103/PhysRevLett.122.090401} {\bibfield  {journal}
  {\bibinfo  {journal} {Physical Review Letters}\ }\textbf {\bibinfo {volume}
  {122}},\ \bibinfo {pages} {090401} (\bibinfo {year} {2019})}\BibitemShut
  {NoStop}%
\bibitem [{\citenamefont {Flynn}\ \emph {et~al.}(2022)\citenamefont {Flynn},
  \citenamefont {Parisi}, \citenamefont {Billam},\ and\ \citenamefont
  {Parker}}]{flynn_2022_quantum}%
  \BibitemOpen
  \bibfield  {author} {\bibinfo {author} {\bibfnamefont {T.~A.}\ \bibnamefont
  {Flynn}}, \bibinfo {author} {\bibfnamefont {L.}~\bibnamefont {Parisi}},
  \bibinfo {author} {\bibfnamefont {T.~P.}\ \bibnamefont {Billam}},\ and\
  \bibinfo {author} {\bibfnamefont {N.~G.}\ \bibnamefont {Parker}},\
  }\href@noop {} {\bibinfo {title} {Quantum {{Droplets}} in {{Imbalanced Atomic
  Mixtures}}}} (\bibinfo {year} {2022}),\ \Eprint
  {https://arxiv.org/abs/2209.04318} {arXiv:2209.04318 [cond-mat]} \BibitemShut
  {NoStop}%
\bibitem [{\citenamefont {Guo}\ \emph {et~al.}(2021)\citenamefont {Guo},
  \citenamefont {Jia}, \citenamefont {Li}, \citenamefont {Ma}, \citenamefont
  {Hutson}, \citenamefont {Cui},\ and\ \citenamefont
  {Wang}}]{guo_2021_leehuangyang}%
  \BibitemOpen
  \bibfield  {author} {\bibinfo {author} {\bibfnamefont {Z.}~\bibnamefont
  {Guo}}, \bibinfo {author} {\bibfnamefont {F.}~\bibnamefont {Jia}}, \bibinfo
  {author} {\bibfnamefont {L.}~\bibnamefont {Li}}, \bibinfo {author}
  {\bibfnamefont {Y.}~\bibnamefont {Ma}}, \bibinfo {author} {\bibfnamefont
  {J.~M.}\ \bibnamefont {Hutson}}, \bibinfo {author} {\bibfnamefont
  {X.}~\bibnamefont {Cui}},\ and\ \bibinfo {author} {\bibfnamefont
  {D.}~\bibnamefont {Wang}},\ }\href
  {https://doi.org/10.1103/PhysRevResearch.3.033247} {\bibfield  {journal}
  {\bibinfo  {journal} {Physical Review Research}\ }\textbf {\bibinfo {volume}
  {3}},\ \bibinfo {pages} {033247} (\bibinfo {year} {2021})}\BibitemShut
  {NoStop}%
\bibitem [{\citenamefont {D'Errico}\ \emph {et~al.}(2019)\citenamefont
  {D'Errico}, \citenamefont {Burchianti}, \citenamefont {Prevedelli},
  \citenamefont {Salasnich}, \citenamefont {Ancilotto}, \citenamefont
  {Modugno}, \citenamefont {Minardi},\ and\ \citenamefont
  {Fort}}]{derrico_2019_observation}%
  \BibitemOpen
  \bibfield  {author} {\bibinfo {author} {\bibfnamefont {C.}~\bibnamefont
  {D'Errico}}, \bibinfo {author} {\bibfnamefont {A.}~\bibnamefont
  {Burchianti}}, \bibinfo {author} {\bibfnamefont {M.}~\bibnamefont
  {Prevedelli}}, \bibinfo {author} {\bibfnamefont {L.}~\bibnamefont
  {Salasnich}}, \bibinfo {author} {\bibfnamefont {F.}~\bibnamefont
  {Ancilotto}}, \bibinfo {author} {\bibfnamefont {M.}~\bibnamefont {Modugno}},
  \bibinfo {author} {\bibfnamefont {F.}~\bibnamefont {Minardi}},\ and\ \bibinfo
  {author} {\bibfnamefont {C.}~\bibnamefont {Fort}},\ }\href
  {https://doi.org/10.1103/PhysRevResearch.1.033155} {\bibfield  {journal}
  {\bibinfo  {journal} {Physical Review Research}\ }\textbf {\bibinfo {volume}
  {1}},\ \bibinfo {pages} {033155} (\bibinfo {year} {2019})}\BibitemShut
  {NoStop}%
\bibitem [{\citenamefont {Zhang}\ \emph {et~al.}(2016)\citenamefont {Zhang},
  \citenamefont {Wen}, \citenamefont {Dai}, \citenamefont {Dong}, \citenamefont
  {Jiang}, \citenamefont {Chang},\ and\ \citenamefont
  {Zhang}}]{zhang_2016_exotic}%
  \BibitemOpen
  \bibfield  {author} {\bibinfo {author} {\bibfnamefont {X.-F.}\ \bibnamefont
  {Zhang}}, \bibinfo {author} {\bibfnamefont {L.}~\bibnamefont {Wen}}, \bibinfo
  {author} {\bibfnamefont {C.-Q.}\ \bibnamefont {Dai}}, \bibinfo {author}
  {\bibfnamefont {R.-F.}\ \bibnamefont {Dong}}, \bibinfo {author}
  {\bibfnamefont {H.-F.}\ \bibnamefont {Jiang}}, \bibinfo {author}
  {\bibfnamefont {H.}~\bibnamefont {Chang}},\ and\ \bibinfo {author}
  {\bibfnamefont {S.-G.}\ \bibnamefont {Zhang}},\ }\href
  {https://doi.org/10.1038/srep19380} {\bibfield  {journal} {\bibinfo
  {journal} {Scientific Reports}\ }\textbf {\bibinfo {volume} {6}},\ \bibinfo
  {pages} {19380} (\bibinfo {year} {2016})}\BibitemShut {NoStop}%
\bibitem [{\citenamefont {Roccuzzo}\ \emph {et~al.}(2020)\citenamefont
  {Roccuzzo}, \citenamefont {Gallem{\'i}}, \citenamefont {Recati},\ and\
  \citenamefont {Stringari}}]{roccuzzo_2020_rotating}%
  \BibitemOpen
  \bibfield  {author} {\bibinfo {author} {\bibfnamefont {S.~M.}\ \bibnamefont
  {Roccuzzo}}, \bibinfo {author} {\bibfnamefont {A.}~\bibnamefont
  {Gallem{\'i}}}, \bibinfo {author} {\bibfnamefont {A.}~\bibnamefont
  {Recati}},\ and\ \bibinfo {author} {\bibfnamefont {S.}~\bibnamefont
  {Stringari}},\ }\href {https://doi.org/10.1103/PhysRevLett.124.045702}
  {\bibfield  {journal} {\bibinfo  {journal} {Physical Review Letters}\
  }\textbf {\bibinfo {volume} {124}},\ \bibinfo {pages} {045702} (\bibinfo
  {year} {2020})}\BibitemShut {NoStop}%
\bibitem [{\citenamefont {Gallem{\`i}}\ \emph {et~al.}(2020)\citenamefont
  {Gallem{\`i}}, \citenamefont {Roccuzzo}, \citenamefont {Stringari},\ and\
  \citenamefont {Recati}}]{gallemi_2020_quantized}%
  \BibitemOpen
  \bibfield  {author} {\bibinfo {author} {\bibfnamefont {A.}~\bibnamefont
  {Gallem{\`i}}}, \bibinfo {author} {\bibfnamefont {S.~M.}\ \bibnamefont
  {Roccuzzo}}, \bibinfo {author} {\bibfnamefont {S.}~\bibnamefont
  {Stringari}},\ and\ \bibinfo {author} {\bibfnamefont {A.}~\bibnamefont
  {Recati}},\ }\href {https://doi.org/10.1103/PhysRevA.102.023322} {\bibfield
  {journal} {\bibinfo  {journal} {Physical Review A}\ }\textbf {\bibinfo
  {volume} {102}},\ \bibinfo {pages} {023322} (\bibinfo {year}
  {2020})}\BibitemShut {NoStop}%
\bibitem [{\citenamefont {Klaus}\ \emph {et~al.}(2022)\citenamefont {Klaus},
  \citenamefont {Bland}, \citenamefont {Poli}, \citenamefont {Politi},
  \citenamefont {Lamporesi}, \citenamefont {Casotti}, \citenamefont {Bisset},
  \citenamefont {Mark},\ and\ \citenamefont
  {Ferlaino}}]{klaus_2022_observation}%
  \BibitemOpen
  \bibfield  {author} {\bibinfo {author} {\bibfnamefont {L.}~\bibnamefont
  {Klaus}}, \bibinfo {author} {\bibfnamefont {T.}~\bibnamefont {Bland}},
  \bibinfo {author} {\bibfnamefont {E.}~\bibnamefont {Poli}}, \bibinfo {author}
  {\bibfnamefont {C.}~\bibnamefont {Politi}}, \bibinfo {author} {\bibfnamefont
  {G.}~\bibnamefont {Lamporesi}}, \bibinfo {author} {\bibfnamefont
  {E.}~\bibnamefont {Casotti}}, \bibinfo {author} {\bibfnamefont {R.~N.}\
  \bibnamefont {Bisset}}, \bibinfo {author} {\bibfnamefont {M.~J.}\
  \bibnamefont {Mark}},\ and\ \bibinfo {author} {\bibfnamefont
  {F.}~\bibnamefont {Ferlaino}},\ }\href
  {https://doi.org/10.48550/arXiv.2206.12265} {\bibinfo {title} {Observation of
  vortices and vortex stripes in a dipolar {{Bose-Einstein}} condensate}}
  (\bibinfo {year} {2022}),\ \Eprint {https://arxiv.org/abs/2206.12265}
  {arXiv:2206.12265 [cond-mat]} \BibitemShut {NoStop}%
\bibitem [{\citenamefont {Prasad}\ \emph {et~al.}(2019)\citenamefont {Prasad},
  \citenamefont {Bland}, \citenamefont {Mulkerin}, \citenamefont {Parker},\
  and\ \citenamefont {Martin}}]{prasad_2019_instability}%
  \BibitemOpen
  \bibfield  {author} {\bibinfo {author} {\bibfnamefont {S.~B.}\ \bibnamefont
  {Prasad}}, \bibinfo {author} {\bibfnamefont {T.}~\bibnamefont {Bland}},
  \bibinfo {author} {\bibfnamefont {B.~C.}\ \bibnamefont {Mulkerin}}, \bibinfo
  {author} {\bibfnamefont {N.~G.}\ \bibnamefont {Parker}},\ and\ \bibinfo
  {author} {\bibfnamefont {A.~M.}\ \bibnamefont {Martin}},\ }\href
  {https://doi.org/10.1103/PhysRevLett.122.050401} {\bibfield  {journal}
  {\bibinfo  {journal} {Physical Review Letters}\ }\textbf {\bibinfo {volume}
  {122}},\ \bibinfo {pages} {050401} (\bibinfo {year} {2019})}\BibitemShut
  {NoStop}%
\bibitem [{\citenamefont {Baillie}\ and\ \citenamefont
  {Blakie}(2020)}]{baillie_2020_rotational}%
  \BibitemOpen
  \bibfield  {author} {\bibinfo {author} {\bibfnamefont {D.}~\bibnamefont
  {Baillie}}\ and\ \bibinfo {author} {\bibfnamefont {P.~B.}\ \bibnamefont
  {Blakie}},\ }\href {https://doi.org/10.1103/PhysRevA.101.043606} {\bibfield
  {journal} {\bibinfo  {journal} {Physical Review A}\ }\textbf {\bibinfo
  {volume} {101}},\ \bibinfo {pages} {043606} (\bibinfo {year}
  {2020})}\BibitemShut {NoStop}%
\bibitem [{\citenamefont {Prasad}\ \emph {et~al.}(2021)\citenamefont {Prasad},
  \citenamefont {Mulkerin},\ and\ \citenamefont
  {Martin}}]{prasad_2021_arbitraryangle}%
  \BibitemOpen
  \bibfield  {author} {\bibinfo {author} {\bibfnamefont {S.~B.}\ \bibnamefont
  {Prasad}}, \bibinfo {author} {\bibfnamefont {B.~C.}\ \bibnamefont
  {Mulkerin}},\ and\ \bibinfo {author} {\bibfnamefont {A.~M.}\ \bibnamefont
  {Martin}},\ }\href {https://doi.org/10.1103/PhysRevA.103.033322} {\bibfield
  {journal} {\bibinfo  {journal} {Physical Review A}\ }\textbf {\bibinfo
  {volume} {103}},\ \bibinfo {pages} {033322} (\bibinfo {year}
  {2021})}\BibitemShut {NoStop}%
\bibitem [{\citenamefont {Halder}\ \emph {et~al.}(2022)\citenamefont {Halder},
  \citenamefont {Mukherjee}, \citenamefont {Mistakidis}, \citenamefont {Das},
  \citenamefont {Kevrekidis}, \citenamefont {Panigrahi}, \citenamefont
  {Majumder},\ and\ \citenamefont {Sadeghpour}}]{halder_2022_control}%
  \BibitemOpen
  \bibfield  {author} {\bibinfo {author} {\bibfnamefont {S.}~\bibnamefont
  {Halder}}, \bibinfo {author} {\bibfnamefont {K.}~\bibnamefont {Mukherjee}},
  \bibinfo {author} {\bibfnamefont {S.~I.}\ \bibnamefont {Mistakidis}},
  \bibinfo {author} {\bibfnamefont {S.}~\bibnamefont {Das}}, \bibinfo {author}
  {\bibfnamefont {P.~G.}\ \bibnamefont {Kevrekidis}}, \bibinfo {author}
  {\bibfnamefont {P.~K.}\ \bibnamefont {Panigrahi}}, \bibinfo {author}
  {\bibfnamefont {S.}~\bibnamefont {Majumder}},\ and\ \bibinfo {author}
  {\bibfnamefont {H.~R.}\ \bibnamefont {Sadeghpour}},\ }\href
  {https://doi.org/10.1103/PhysRevResearch.4.043124} {\bibfield  {journal}
  {\bibinfo  {journal} {Phys. Rev. Res.}\ }\textbf {\bibinfo {volume} {4}},\
  \bibinfo {pages} {043124} (\bibinfo {year} {2022})}\BibitemShut {NoStop}%
\bibitem [{\citenamefont {Zhang}\ \emph {et~al.}(2022)\citenamefont {Zhang},
  \citenamefont {Zhang}, \citenamefont {Yang},\ and\ \citenamefont
  {{Capogrosso-Sansone}}}]{zhang_2022_supersolid}%
  \BibitemOpen
  \bibfield  {author} {\bibinfo {author} {\bibfnamefont {J.}~\bibnamefont
  {Zhang}}, \bibinfo {author} {\bibfnamefont {C.}~\bibnamefont {Zhang}},
  \bibinfo {author} {\bibfnamefont {J.}~\bibnamefont {Yang}},\ and\ \bibinfo
  {author} {\bibfnamefont {B.}~\bibnamefont {{Capogrosso-Sansone}}},\ }\href
  {https://doi.org/10.1103/PhysRevA.105.063302} {\bibfield  {journal} {\bibinfo
   {journal} {Physical Review A}\ }\textbf {\bibinfo {volume} {105}},\ \bibinfo
  {pages} {063302} (\bibinfo {year} {2022})}\BibitemShut {NoStop}%
\bibitem [{\citenamefont {Heidarian}\ and\ \citenamefont
  {Paramekanti}(2010)}]{heidarian_2010_supersolidity}%
  \BibitemOpen
  \bibfield  {author} {\bibinfo {author} {\bibfnamefont {D.}~\bibnamefont
  {Heidarian}}\ and\ \bibinfo {author} {\bibfnamefont {A.}~\bibnamefont
  {Paramekanti}},\ }\href {https://doi.org/10.1103/PhysRevLett.104.015301}
  {\bibfield  {journal} {\bibinfo  {journal} {Physical Review Letters}\
  }\textbf {\bibinfo {volume} {104}},\ \bibinfo {pages} {015301} (\bibinfo
  {year} {2010})}\BibitemShut {NoStop}%
\bibitem [{\citenamefont {Suthar}\ \emph {et~al.}(2020)\citenamefont {Suthar},
  \citenamefont {Sable}, \citenamefont {Bai}, \citenamefont {Bandyopadhyay},
  \citenamefont {Pal},\ and\ \citenamefont {Angom}}]{suthar_2020_supersolid}%
  \BibitemOpen
  \bibfield  {author} {\bibinfo {author} {\bibfnamefont {K.}~\bibnamefont
  {Suthar}}, \bibinfo {author} {\bibfnamefont {H.}~\bibnamefont {Sable}},
  \bibinfo {author} {\bibfnamefont {R.}~\bibnamefont {Bai}}, \bibinfo {author}
  {\bibfnamefont {S.}~\bibnamefont {Bandyopadhyay}}, \bibinfo {author}
  {\bibfnamefont {S.}~\bibnamefont {Pal}},\ and\ \bibinfo {author}
  {\bibfnamefont {D.}~\bibnamefont {Angom}},\ }\href
  {https://doi.org/10.1103/PhysRevA.102.013320} {\bibfield  {journal} {\bibinfo
   {journal} {Physical Review A}\ }\textbf {\bibinfo {volume} {102}},\ \bibinfo
  {pages} {013320} (\bibinfo {year} {2020})}\BibitemShut {NoStop}%
\bibitem [{\citenamefont {H{\'e}bert}\ \emph {et~al.}(2008)\citenamefont
  {H{\'e}bert}, \citenamefont {Batrouni}, \citenamefont {Roy},\ and\
  \citenamefont {Rousseau}}]{hebert_2008_supersolids}%
  \BibitemOpen
  \bibfield  {author} {\bibinfo {author} {\bibfnamefont {F.}~\bibnamefont
  {H{\'e}bert}}, \bibinfo {author} {\bibfnamefont {G.~G.}\ \bibnamefont
  {Batrouni}}, \bibinfo {author} {\bibfnamefont {X.}~\bibnamefont {Roy}},\ and\
  \bibinfo {author} {\bibfnamefont {V.~G.}\ \bibnamefont {Rousseau}},\ }\href
  {https://doi.org/10.1103/PhysRevB.78.184505} {\bibfield  {journal} {\bibinfo
  {journal} {Physical Review B}\ }\textbf {\bibinfo {volume} {78}},\ \bibinfo
  {pages} {184505} (\bibinfo {year} {2008})}\BibitemShut {NoStop}%
\bibitem [{\citenamefont {Cinti}\ \emph {et~al.}(2010)\citenamefont {Cinti},
  \citenamefont {Jain}, \citenamefont {Boninsegni}, \citenamefont {Micheli},
  \citenamefont {Zoller},\ and\ \citenamefont
  {Pupillo}}]{cinti_2010_supersolid}%
  \BibitemOpen
  \bibfield  {author} {\bibinfo {author} {\bibfnamefont {F.}~\bibnamefont
  {Cinti}}, \bibinfo {author} {\bibfnamefont {P.}~\bibnamefont {Jain}},
  \bibinfo {author} {\bibfnamefont {M.}~\bibnamefont {Boninsegni}}, \bibinfo
  {author} {\bibfnamefont {A.}~\bibnamefont {Micheli}}, \bibinfo {author}
  {\bibfnamefont {P.}~\bibnamefont {Zoller}},\ and\ \bibinfo {author}
  {\bibfnamefont {G.}~\bibnamefont {Pupillo}},\ }\href
  {https://doi.org/10.1103/PhysRevLett.105.135301} {\bibfield  {journal}
  {\bibinfo  {journal} {Physical Review Letters}\ }\textbf {\bibinfo {volume}
  {105}},\ \bibinfo {pages} {135301} (\bibinfo {year} {2010})}\BibitemShut
  {NoStop}%
\bibitem [{\citenamefont {Henkel}\ \emph {et~al.}(2012)\citenamefont {Henkel},
  \citenamefont {Cinti}, \citenamefont {Jain}, \citenamefont {Pupillo},\ and\
  \citenamefont {Pohl}}]{henkel_2012_supersolid}%
  \BibitemOpen
  \bibfield  {author} {\bibinfo {author} {\bibfnamefont {N.}~\bibnamefont
  {Henkel}}, \bibinfo {author} {\bibfnamefont {F.}~\bibnamefont {Cinti}},
  \bibinfo {author} {\bibfnamefont {P.}~\bibnamefont {Jain}}, \bibinfo {author}
  {\bibfnamefont {G.}~\bibnamefont {Pupillo}},\ and\ \bibinfo {author}
  {\bibfnamefont {T.}~\bibnamefont {Pohl}},\ }\href
  {https://doi.org/10.1103/PhysRevLett.108.265301} {\bibfield  {journal}
  {\bibinfo  {journal} {Physical Review Letters}\ }\textbf {\bibinfo {volume}
  {108}},\ \bibinfo {pages} {265301} (\bibinfo {year} {2012})}\BibitemShut
  {NoStop}%
\bibitem [{\citenamefont {Li}\ \emph {et~al.}(2017)\citenamefont {Li},
  \citenamefont {Lee}, \citenamefont {Huang}, \citenamefont {Burchesky},
  \citenamefont {Shteynas}, \citenamefont {Top}, \citenamefont {Jamison},\ and\
  \citenamefont {Ketterle}}]{li_2017_stripe}%
  \BibitemOpen
  \bibfield  {author} {\bibinfo {author} {\bibfnamefont {J.-R.}\ \bibnamefont
  {Li}}, \bibinfo {author} {\bibfnamefont {J.}~\bibnamefont {Lee}}, \bibinfo
  {author} {\bibfnamefont {W.}~\bibnamefont {Huang}}, \bibinfo {author}
  {\bibfnamefont {S.}~\bibnamefont {Burchesky}}, \bibinfo {author}
  {\bibfnamefont {B.}~\bibnamefont {Shteynas}}, \bibinfo {author}
  {\bibfnamefont {F.~{\c C}.}\ \bibnamefont {Top}}, \bibinfo {author}
  {\bibfnamefont {A.~O.}\ \bibnamefont {Jamison}},\ and\ \bibinfo {author}
  {\bibfnamefont {W.}~\bibnamefont {Ketterle}},\ }\href
  {https://doi.org/10.1038/nature21431} {\bibfield  {journal} {\bibinfo
  {journal} {Nature}\ }\textbf {\bibinfo {volume} {543}},\ \bibinfo {pages}
  {91} (\bibinfo {year} {2017})}\BibitemShut {NoStop}%
\bibitem [{\citenamefont {Sachdeva}\ \emph {et~al.}(2020)\citenamefont
  {Sachdeva}, \citenamefont {Tengstrand},\ and\ \citenamefont
  {Reimann}}]{sachdeva_2020_selfbound}%
  \BibitemOpen
  \bibfield  {author} {\bibinfo {author} {\bibfnamefont {R.}~\bibnamefont
  {Sachdeva}}, \bibinfo {author} {\bibfnamefont {M.~N.}\ \bibnamefont
  {Tengstrand}},\ and\ \bibinfo {author} {\bibfnamefont {S.~M.}\ \bibnamefont
  {Reimann}},\ }\href {https://doi.org/10.1103/PhysRevA.102.043304} {\bibfield
  {journal} {\bibinfo  {journal} {Physical Review A}\ }\textbf {\bibinfo
  {volume} {102}},\ \bibinfo {pages} {043304} (\bibinfo {year}
  {2020})}\BibitemShut {NoStop}%
\bibitem [{\citenamefont {Schmidt}\ \emph {et~al.}(2022)\citenamefont
  {Schmidt}, \citenamefont {Lassabli{\`e}re}, \citenamefont
  {Qu{\'e}m{\'e}ner},\ and\ \citenamefont {Langen}}]{schmidt_2022_selfbound}%
  \BibitemOpen
  \bibfield  {author} {\bibinfo {author} {\bibfnamefont {M.}~\bibnamefont
  {Schmidt}}, \bibinfo {author} {\bibfnamefont {L.}~\bibnamefont
  {Lassabli{\`e}re}}, \bibinfo {author} {\bibfnamefont {G.}~\bibnamefont
  {Qu{\'e}m{\'e}ner}},\ and\ \bibinfo {author} {\bibfnamefont {T.}~\bibnamefont
  {Langen}},\ }\href {https://doi.org/10.1103/PhysRevResearch.4.013235}
  {\bibfield  {journal} {\bibinfo  {journal} {Physical Review Research}\
  }\textbf {\bibinfo {volume} {4}},\ \bibinfo {pages} {013235} (\bibinfo {year}
  {2022})}\BibitemShut {NoStop}%
\bibitem [{\citenamefont {Li}\ \emph {et~al.}(2022)\citenamefont {Li},
  \citenamefont {Le},\ and\ \citenamefont {Saito}}]{li_2022_longlifetime}%
  \BibitemOpen
  \bibfield  {author} {\bibinfo {author} {\bibfnamefont {S.}~\bibnamefont
  {Li}}, \bibinfo {author} {\bibfnamefont {U.~N.}\ \bibnamefont {Le}},\ and\
  \bibinfo {author} {\bibfnamefont {H.}~\bibnamefont {Saito}},\ }\href
  {https://doi.org/10.1103/PhysRevA.105.L061302} {\bibfield  {journal}
  {\bibinfo  {journal} {Physical Review A}\ }\textbf {\bibinfo {volume}
  {105}},\ \bibinfo {pages} {L061302} (\bibinfo {year} {2022})}\BibitemShut
  {NoStop}%
\bibitem [{\citenamefont {Bland}\ \emph
  {et~al.}(2022{\natexlab{b}})\citenamefont {Bland}, \citenamefont {Poli},
  \citenamefont {Ardila}, \citenamefont {Santos}, \citenamefont {Ferlaino},\
  and\ \citenamefont {Bisset}}]{bland_2022_alternatingdomain}%
  \BibitemOpen
  \bibfield  {author} {\bibinfo {author} {\bibfnamefont {T.}~\bibnamefont
  {Bland}}, \bibinfo {author} {\bibfnamefont {E.}~\bibnamefont {Poli}},
  \bibinfo {author} {\bibfnamefont {L.~A. P.~n.}\ \bibnamefont {Ardila}},
  \bibinfo {author} {\bibfnamefont {L.}~\bibnamefont {Santos}}, \bibinfo
  {author} {\bibfnamefont {F.}~\bibnamefont {Ferlaino}},\ and\ \bibinfo
  {author} {\bibfnamefont {R.~N.}\ \bibnamefont {Bisset}},\ }\href
  {https://doi.org/10.1103/PhysRevA.106.053322} {\bibfield  {journal} {\bibinfo
   {journal} {Phys. Rev. A}\ }\textbf {\bibinfo {volume} {106}},\ \bibinfo
  {pages} {053322} (\bibinfo {year} {2022}{\natexlab{b}})}\BibitemShut
  {NoStop}%
\bibitem [{\citenamefont {Trautmann}\ \emph {et~al.}(2018)\citenamefont
  {Trautmann}, \citenamefont {Ilzh{\"o}fer}, \citenamefont {Durastante},
  \citenamefont {Politi}, \citenamefont {Sohmen}, \citenamefont {Mark},\ and\
  \citenamefont {Ferlaino}}]{trautmann_2018_dipolar}%
  \BibitemOpen
  \bibfield  {author} {\bibinfo {author} {\bibfnamefont {A.}~\bibnamefont
  {Trautmann}}, \bibinfo {author} {\bibfnamefont {P.}~\bibnamefont
  {Ilzh{\"o}fer}}, \bibinfo {author} {\bibfnamefont {G.}~\bibnamefont
  {Durastante}}, \bibinfo {author} {\bibfnamefont {C.}~\bibnamefont {Politi}},
  \bibinfo {author} {\bibfnamefont {M.}~\bibnamefont {Sohmen}}, \bibinfo
  {author} {\bibfnamefont {M.~J.}\ \bibnamefont {Mark}},\ and\ \bibinfo
  {author} {\bibfnamefont {F.}~\bibnamefont {Ferlaino}},\ }\href
  {https://doi.org/10.1103/PhysRevLett.121.213601} {\bibfield  {journal}
  {\bibinfo  {journal} {Physical Review Letters}\ }\textbf {\bibinfo {volume}
  {121}},\ \bibinfo {pages} {213601} (\bibinfo {year} {2018})}\BibitemShut
  {NoStop}%
\bibitem [{\citenamefont {Durastante}\ \emph {et~al.}(2020)\citenamefont
  {Durastante}, \citenamefont {Politi}, \citenamefont {Sohmen}, \citenamefont
  {Ilzh{\"o}fer}, \citenamefont {Mark}, \citenamefont {Norcia},\ and\
  \citenamefont {Ferlaino}}]{durastante_2020_feshbach}%
  \BibitemOpen
  \bibfield  {author} {\bibinfo {author} {\bibfnamefont {G.}~\bibnamefont
  {Durastante}}, \bibinfo {author} {\bibfnamefont {C.}~\bibnamefont {Politi}},
  \bibinfo {author} {\bibfnamefont {M.}~\bibnamefont {Sohmen}}, \bibinfo
  {author} {\bibfnamefont {P.}~\bibnamefont {Ilzh{\"o}fer}}, \bibinfo {author}
  {\bibfnamefont {M.~J.}\ \bibnamefont {Mark}}, \bibinfo {author}
  {\bibfnamefont {M.~A.}\ \bibnamefont {Norcia}},\ and\ \bibinfo {author}
  {\bibfnamefont {F.}~\bibnamefont {Ferlaino}},\ }\href
  {https://doi.org/10.1103/PhysRevA.102.033330} {\bibfield  {journal} {\bibinfo
   {journal} {Physical Review A}\ }\textbf {\bibinfo {volume} {102}},\ \bibinfo
  {pages} {033330} (\bibinfo {year} {2020})}\BibitemShut {NoStop}%
\bibitem [{\citenamefont {Politi}\ \emph {et~al.}(2022)\citenamefont {Politi},
  \citenamefont {Trautmann}, \citenamefont {Ilzh{\"o}fer}, \citenamefont
  {Durastante}, \citenamefont {Mark}, \citenamefont {Modugno},\ and\
  \citenamefont {Ferlaino}}]{politi_2022_interspecies}%
  \BibitemOpen
  \bibfield  {author} {\bibinfo {author} {\bibfnamefont {C.}~\bibnamefont
  {Politi}}, \bibinfo {author} {\bibfnamefont {A.}~\bibnamefont {Trautmann}},
  \bibinfo {author} {\bibfnamefont {P.}~\bibnamefont {Ilzh{\"o}fer}}, \bibinfo
  {author} {\bibfnamefont {G.}~\bibnamefont {Durastante}}, \bibinfo {author}
  {\bibfnamefont {M.~J.}\ \bibnamefont {Mark}}, \bibinfo {author}
  {\bibfnamefont {M.}~\bibnamefont {Modugno}},\ and\ \bibinfo {author}
  {\bibfnamefont {F.}~\bibnamefont {Ferlaino}},\ }\href
  {https://doi.org/10.1103/PhysRevA.105.023304} {\bibfield  {journal} {\bibinfo
   {journal} {Physical Review A}\ }\textbf {\bibinfo {volume} {105}},\ \bibinfo
  {pages} {023304} (\bibinfo {year} {2022})}\BibitemShut {NoStop}%
\bibitem [{\citenamefont {Bisset}\ \emph {et~al.}(2021)\citenamefont {Bisset},
  \citenamefont {Ardila},\ and\ \citenamefont {Santos}}]{bisset_2021_quantum}%
  \BibitemOpen
  \bibfield  {author} {\bibinfo {author} {\bibfnamefont {R.~N.}\ \bibnamefont
  {Bisset}}, \bibinfo {author} {\bibfnamefont {L.~A.~P.}\ \bibnamefont
  {Ardila}},\ and\ \bibinfo {author} {\bibfnamefont {L.}~\bibnamefont
  {Santos}},\ }\href {https://doi.org/10.1103/PhysRevLett.126.025301}
  {\bibfield  {journal} {\bibinfo  {journal} {Physical Review Letters}\
  }\textbf {\bibinfo {volume} {126}},\ \bibinfo {pages} {025301} (\bibinfo
  {year} {2021})}\BibitemShut {NoStop}%
\bibitem [{\citenamefont {Smith}\ \emph
  {et~al.}(2021{\natexlab{a}})\citenamefont {Smith}, \citenamefont {Blakie},\
  and\ \citenamefont {Baillie}}]{smith_2021_approximate}%
  \BibitemOpen
  \bibfield  {author} {\bibinfo {author} {\bibfnamefont {J.~C.}\ \bibnamefont
  {Smith}}, \bibinfo {author} {\bibfnamefont {P.~B.}\ \bibnamefont {Blakie}},\
  and\ \bibinfo {author} {\bibfnamefont {D.}~\bibnamefont {Baillie}},\ }\href
  {https://doi.org/10.1103/PhysRevA.104.053316} {\bibfield  {journal} {\bibinfo
   {journal} {Physical Review A}\ }\textbf {\bibinfo {volume} {104}},\ \bibinfo
  {pages} {053316} (\bibinfo {year} {2021}{\natexlab{a}})}\BibitemShut
  {NoStop}%
\bibitem [{\citenamefont {Lee}\ \emph {et~al.}(2022)\citenamefont {Lee},
  \citenamefont {Baillie},\ and\ \citenamefont {Blakie}}]{lee_2022_stability}%
  \BibitemOpen
  \bibfield  {author} {\bibinfo {author} {\bibfnamefont {A.-C.}\ \bibnamefont
  {Lee}}, \bibinfo {author} {\bibfnamefont {D.}~\bibnamefont {Baillie}},\ and\
  \bibinfo {author} {\bibfnamefont {P.~B.}\ \bibnamefont {Blakie}},\ }\href
  {https://doi.org/10.1103/PhysRevResearch.4.033153} {\bibfield  {journal}
  {\bibinfo  {journal} {Physical Review Research}\ }\textbf {\bibinfo {volume}
  {4}},\ \bibinfo {pages} {033153} (\bibinfo {year} {2022})}\BibitemShut
  {NoStop}%
\bibitem [{\citenamefont {Lee}\ \emph {et~al.}(2021{\natexlab{b}})\citenamefont
  {Lee}, \citenamefont {Baillie}, \citenamefont {Blakie},\ and\ \citenamefont
  {Bisset}}]{lee_2021_miscibility}%
  \BibitemOpen
  \bibfield  {author} {\bibinfo {author} {\bibfnamefont {A.-C.}\ \bibnamefont
  {Lee}}, \bibinfo {author} {\bibfnamefont {D.}~\bibnamefont {Baillie}},
  \bibinfo {author} {\bibfnamefont {P.~B.}\ \bibnamefont {Blakie}},\ and\
  \bibinfo {author} {\bibfnamefont {R.~N.}\ \bibnamefont {Bisset}},\ }\href
  {https://doi.org/10.1103/PhysRevA.103.063301} {\bibfield  {journal} {\bibinfo
   {journal} {Physical Review A}\ }\textbf {\bibinfo {volume} {103}},\ \bibinfo
  {pages} {063301} (\bibinfo {year} {2021}{\natexlab{b}})}\BibitemShut
  {NoStop}%
\bibitem [{\citenamefont {Smith}\ \emph
  {et~al.}(2021{\natexlab{b}})\citenamefont {Smith}, \citenamefont {Baillie},\
  and\ \citenamefont {Blakie}}]{smith_2021_quantum}%
  \BibitemOpen
  \bibfield  {author} {\bibinfo {author} {\bibfnamefont {J.~C.}\ \bibnamefont
  {Smith}}, \bibinfo {author} {\bibfnamefont {D.}~\bibnamefont {Baillie}},\
  and\ \bibinfo {author} {\bibfnamefont {P.~B.}\ \bibnamefont {Blakie}},\
  }\href {https://doi.org/10.1103/PhysRevLett.126.025302} {\bibfield  {journal}
  {\bibinfo  {journal} {Physical Review Letters}\ }\textbf {\bibinfo {volume}
  {126}},\ \bibinfo {pages} {025302} (\bibinfo {year}
  {2021}{\natexlab{b}})}\BibitemShut {NoStop}%
\bibitem [{\citenamefont {Scheiermann}\ \emph {et~al.}(2022)\citenamefont
  {Scheiermann}, \citenamefont {Ardila}, \citenamefont {Bland}, \citenamefont
  {Bisset},\ and\ \citenamefont {Santos}}]{scheiermann_2022_catalyzation}%
  \BibitemOpen
  \bibfield  {author} {\bibinfo {author} {\bibfnamefont {D.}~\bibnamefont
  {Scheiermann}}, \bibinfo {author} {\bibfnamefont {L.~A.~P.}\ \bibnamefont
  {Ardila}}, \bibinfo {author} {\bibfnamefont {T.}~\bibnamefont {Bland}},
  \bibinfo {author} {\bibfnamefont {R.~N.}\ \bibnamefont {Bisset}},\ and\
  \bibinfo {author} {\bibfnamefont {L.}~\bibnamefont {Santos}},\ }\href
  {https://doi.org/10.48550/arXiv.2202.08259} {\bibinfo {title} {Catalyzation
  of supersolidity in binary dipolar condensates}} (\bibinfo {year} {2022}),\
  \Eprint {https://arxiv.org/abs/2202.08259} {arXiv:2202.08259 [cond-mat]}
  \BibitemShut {NoStop}%
\bibitem [{\citenamefont {Schmidt}\ \emph {et~al.}(2021)\citenamefont
  {Schmidt}, \citenamefont {Hertkorn}, \citenamefont {Guo}, \citenamefont
  {B{\"o}ttcher}, \citenamefont {Schmidt}, \citenamefont {Ng}, \citenamefont
  {Graham}, \citenamefont {Langen}, \citenamefont {Zwierlein},\ and\
  \citenamefont {Pfau}}]{schmidt_2021_roton}%
  \BibitemOpen
  \bibfield  {author} {\bibinfo {author} {\bibfnamefont {J.-N.}\ \bibnamefont
  {Schmidt}}, \bibinfo {author} {\bibfnamefont {J.}~\bibnamefont {Hertkorn}},
  \bibinfo {author} {\bibfnamefont {M.}~\bibnamefont {Guo}}, \bibinfo {author}
  {\bibfnamefont {F.}~\bibnamefont {B{\"o}ttcher}}, \bibinfo {author}
  {\bibfnamefont {M.}~\bibnamefont {Schmidt}}, \bibinfo {author} {\bibfnamefont
  {K.~S.~H.}\ \bibnamefont {Ng}}, \bibinfo {author} {\bibfnamefont {S.~D.}\
  \bibnamefont {Graham}}, \bibinfo {author} {\bibfnamefont {T.}~\bibnamefont
  {Langen}}, \bibinfo {author} {\bibfnamefont {M.}~\bibnamefont {Zwierlein}},\
  and\ \bibinfo {author} {\bibfnamefont {T.}~\bibnamefont {Pfau}},\ }\href
  {https://doi.org/10.1103/PhysRevLett.126.193002} {\bibfield  {journal}
  {\bibinfo  {journal} {Physical Review Letters}\ }\textbf {\bibinfo {volume}
  {126}},\ \bibinfo {pages} {193002} (\bibinfo {year} {2021})}\BibitemShut
  {NoStop}%
\bibitem [{\citenamefont {Hertkorn}\ \emph
  {et~al.}(2021{\natexlab{b}})\citenamefont {Hertkorn}, \citenamefont
  {Schmidt}, \citenamefont {Guo}, \citenamefont {B\"ottcher}, \citenamefont
  {Ng}, \citenamefont {Graham}, \citenamefont {Uerlings}, \citenamefont
  {B\"uchler}, \citenamefont {Langen}, \citenamefont {Zwierlein},\ and\
  \citenamefont {Pfau}}]{hertkorn_2021_supersolidity}%
  \BibitemOpen
  \bibfield  {author} {\bibinfo {author} {\bibfnamefont {J.}~\bibnamefont
  {Hertkorn}}, \bibinfo {author} {\bibfnamefont {J.-N.}\ \bibnamefont
  {Schmidt}}, \bibinfo {author} {\bibfnamefont {M.}~\bibnamefont {Guo}},
  \bibinfo {author} {\bibfnamefont {F.}~\bibnamefont {B\"ottcher}}, \bibinfo
  {author} {\bibfnamefont {K.~S.~H.}\ \bibnamefont {Ng}}, \bibinfo {author}
  {\bibfnamefont {S.~D.}\ \bibnamefont {Graham}}, \bibinfo {author}
  {\bibfnamefont {P.}~\bibnamefont {Uerlings}}, \bibinfo {author}
  {\bibfnamefont {H.~P.}\ \bibnamefont {B\"uchler}}, \bibinfo {author}
  {\bibfnamefont {T.}~\bibnamefont {Langen}}, \bibinfo {author} {\bibfnamefont
  {M.}~\bibnamefont {Zwierlein}},\ and\ \bibinfo {author} {\bibfnamefont
  {T.}~\bibnamefont {Pfau}},\ }\href
  {https://doi.org/10.1103/PhysRevLett.127.155301} {\bibfield  {journal}
  {\bibinfo  {journal} {Phys. Rev. Lett.}\ }\textbf {\bibinfo {volume} {127}},\
  \bibinfo {pages} {155301} (\bibinfo {year} {2021}{\natexlab{b}})}\BibitemShut
  {NoStop}%
\bibitem [{\citenamefont {Chomaz}\ \emph {et~al.}(2018)\citenamefont {Chomaz},
  \citenamefont {{\noopsort{bijnen}}{van Bijnen}}, \citenamefont {Petter},
  \citenamefont {Faraoni}, \citenamefont {Baier}, \citenamefont {Becher},
  \citenamefont {Mark}, \citenamefont {W{\"a}chtler}, \citenamefont {Santos},\
  and\ \citenamefont {Ferlaino}}]{chomaz_2018_observation}%
  \BibitemOpen
  \bibfield  {author} {\bibinfo {author} {\bibfnamefont {L.}~\bibnamefont
  {Chomaz}}, \bibinfo {author} {\bibfnamefont {R.~M.~W.}\ \bibnamefont
  {{\noopsort{bijnen}}{van Bijnen}}}, \bibinfo {author} {\bibfnamefont
  {D.}~\bibnamefont {Petter}}, \bibinfo {author} {\bibfnamefont
  {G.}~\bibnamefont {Faraoni}}, \bibinfo {author} {\bibfnamefont
  {S.}~\bibnamefont {Baier}}, \bibinfo {author} {\bibfnamefont {J.~H.}\
  \bibnamefont {Becher}}, \bibinfo {author} {\bibfnamefont {M.~J.}\
  \bibnamefont {Mark}}, \bibinfo {author} {\bibfnamefont {F.}~\bibnamefont
  {W{\"a}chtler}}, \bibinfo {author} {\bibfnamefont {L.}~\bibnamefont
  {Santos}},\ and\ \bibinfo {author} {\bibfnamefont {F.}~\bibnamefont
  {Ferlaino}},\ }\href {https://doi.org/10.1038/s41567-018-0054-7} {\bibfield
  {journal} {\bibinfo  {journal} {Nature Physics}\ }\textbf {\bibinfo {volume}
  {14}},\ \bibinfo {pages} {442} (\bibinfo {year} {2018})}\BibitemShut
  {NoStop}%
\bibitem [{\citenamefont {{S{\'a}nchez-Baena}}\ \emph
  {et~al.}(2022)\citenamefont {{S{\'a}nchez-Baena}}, \citenamefont {Politi},
  \citenamefont {Maucher}, \citenamefont {Ferlaino},\ and\ \citenamefont
  {Pohl}}]{sanchez-baena_2022_heating}%
  \BibitemOpen
  \bibfield  {author} {\bibinfo {author} {\bibfnamefont {J.}~\bibnamefont
  {{S{\'a}nchez-Baena}}}, \bibinfo {author} {\bibfnamefont {C.}~\bibnamefont
  {Politi}}, \bibinfo {author} {\bibfnamefont {F.}~\bibnamefont {Maucher}},
  \bibinfo {author} {\bibfnamefont {F.}~\bibnamefont {Ferlaino}},\ and\
  \bibinfo {author} {\bibfnamefont {T.}~\bibnamefont {Pohl}},\ }\href
  {https://doi.org/10.48550/arXiv.2209.00335} {\bibinfo {title} {Heating a
  quantum dipolar fluid into a solid}} (\bibinfo {year} {2022}),\ \Eprint
  {https://arxiv.org/abs/2209.00335} {arXiv:2209.00335 [cond-mat]} \BibitemShut
  {NoStop}%
\bibitem [{\citenamefont {De~Rosi}\ \emph {et~al.}(2021)\citenamefont
  {De~Rosi}, \citenamefont {Astrakharchik},\ and\ \citenamefont
  {Massignan}}]{derosi_2021_thermal}%
  \BibitemOpen
  \bibfield  {author} {\bibinfo {author} {\bibfnamefont {G.}~\bibnamefont
  {De~Rosi}}, \bibinfo {author} {\bibfnamefont {G.~E.}\ \bibnamefont
  {Astrakharchik}},\ and\ \bibinfo {author} {\bibfnamefont {P.}~\bibnamefont
  {Massignan}},\ }\href {https://doi.org/10.1103/PhysRevA.103.043316}
  {\bibfield  {journal} {\bibinfo  {journal} {Physical Review A}\ }\textbf
  {\bibinfo {volume} {103}},\ \bibinfo {pages} {043316} (\bibinfo {year}
  {2021})}\BibitemShut {NoStop}%
\bibitem [{\citenamefont {Johnstone}\ \emph {et~al.}(2019)\citenamefont
  {Johnstone}, \citenamefont {Groszek}, \citenamefont {Starkey}, \citenamefont
  {Billington}, \citenamefont {Simula},\ and\ \citenamefont
  {Helmerson}}]{johnstone_2019_evolution}%
  \BibitemOpen
  \bibfield  {author} {\bibinfo {author} {\bibfnamefont {S.~P.}\ \bibnamefont
  {Johnstone}}, \bibinfo {author} {\bibfnamefont {A.~J.}\ \bibnamefont
  {Groszek}}, \bibinfo {author} {\bibfnamefont {P.~T.}\ \bibnamefont
  {Starkey}}, \bibinfo {author} {\bibfnamefont {C.~J.}\ \bibnamefont
  {Billington}}, \bibinfo {author} {\bibfnamefont {T.~P.}\ \bibnamefont
  {Simula}},\ and\ \bibinfo {author} {\bibfnamefont {K.}~\bibnamefont
  {Helmerson}},\ }\href {https://doi.org/10.1126/science.aat5793} {\bibfield
  {journal} {\bibinfo  {journal} {Science}\ }\textbf {\bibinfo {volume}
  {364}},\ \bibinfo {pages} {1267} (\bibinfo {year} {2019})}\BibitemShut
  {NoStop}%
\bibitem [{\citenamefont {Gauthier}\ \emph {et~al.}(2019)\citenamefont
  {Gauthier}, \citenamefont {Reeves}, \citenamefont {Yu}, \citenamefont
  {Bradley}, \citenamefont {Baker}, \citenamefont {Bell}, \citenamefont
  {{Rubinsztein-Dunlop}}, \citenamefont {Davis},\ and\ \citenamefont
  {Neely}}]{gauthier_2019_giant}%
  \BibitemOpen
  \bibfield  {author} {\bibinfo {author} {\bibfnamefont {G.}~\bibnamefont
  {Gauthier}}, \bibinfo {author} {\bibfnamefont {M.~T.}\ \bibnamefont
  {Reeves}}, \bibinfo {author} {\bibfnamefont {X.}~\bibnamefont {Yu}}, \bibinfo
  {author} {\bibfnamefont {A.~S.}\ \bibnamefont {Bradley}}, \bibinfo {author}
  {\bibfnamefont {M.~A.}\ \bibnamefont {Baker}}, \bibinfo {author}
  {\bibfnamefont {T.~A.}\ \bibnamefont {Bell}}, \bibinfo {author}
  {\bibfnamefont {H.}~\bibnamefont {{Rubinsztein-Dunlop}}}, \bibinfo {author}
  {\bibfnamefont {M.~J.}\ \bibnamefont {Davis}},\ and\ \bibinfo {author}
  {\bibfnamefont {T.~W.}\ \bibnamefont {Neely}},\ }\href
  {https://doi.org/10.1126/science.aat5718} {\bibfield  {journal} {\bibinfo
  {journal} {Science}\ }\textbf {\bibinfo {volume} {364}},\ \bibinfo {pages}
  {1264} (\bibinfo {year} {2019})}\BibitemShut {NoStop}%
\bibitem [{\citenamefont {Das}\ \emph {et~al.}(2022)\citenamefont {Das},
  \citenamefont {Mukherjee},\ and\ \citenamefont {Majumder}}]{das_2022_vortex}%
  \BibitemOpen
  \bibfield  {author} {\bibinfo {author} {\bibfnamefont {S.}~\bibnamefont
  {Das}}, \bibinfo {author} {\bibfnamefont {K.}~\bibnamefont {Mukherjee}},\
  and\ \bibinfo {author} {\bibfnamefont {S.}~\bibnamefont {Majumder}},\ }\href
  {https://doi.org/10.1103/PhysRevA.106.023306} {\bibfield  {journal} {\bibinfo
   {journal} {Physical Review A}\ }\textbf {\bibinfo {volume} {106}},\ \bibinfo
  {pages} {023306} (\bibinfo {year} {2022})}\BibitemShut {NoStop}%
\bibitem [{\citenamefont {Zhang}\ \emph {et~al.}(2021)\citenamefont {Zhang},
  \citenamefont {Pohl},\ and\ \citenamefont {Maucher}}]{zhang_2021_phases}%
  \BibitemOpen
  \bibfield  {author} {\bibinfo {author} {\bibfnamefont {Y.-C.}\ \bibnamefont
  {Zhang}}, \bibinfo {author} {\bibfnamefont {T.}~\bibnamefont {Pohl}},\ and\
  \bibinfo {author} {\bibfnamefont {F.}~\bibnamefont {Maucher}},\ }\href
  {https://doi.org/10.1103/PhysRevA.104.013310} {\bibfield  {journal} {\bibinfo
   {journal} {Physical Review A}\ }\textbf {\bibinfo {volume} {104}},\ \bibinfo
  {pages} {013310} (\bibinfo {year} {2021})}\BibitemShut {NoStop}%
\bibitem [{\citenamefont {Maity}\ \emph {et~al.}(2020)\citenamefont {Maity},
  \citenamefont {Mukherjee}, \citenamefont {Mistakidis}, \citenamefont {Das},
  \citenamefont {Kevrekidis}, \citenamefont {Majumder},\ and\ \citenamefont
  {Schmelcher}}]{maity_2020_parametrically}%
  \BibitemOpen
  \bibfield  {author} {\bibinfo {author} {\bibfnamefont {D.~K.}\ \bibnamefont
  {Maity}}, \bibinfo {author} {\bibfnamefont {K.}~\bibnamefont {Mukherjee}},
  \bibinfo {author} {\bibfnamefont {S.~I.}\ \bibnamefont {Mistakidis}},
  \bibinfo {author} {\bibfnamefont {S.}~\bibnamefont {Das}}, \bibinfo {author}
  {\bibfnamefont {P.~G.}\ \bibnamefont {Kevrekidis}}, \bibinfo {author}
  {\bibfnamefont {S.}~\bibnamefont {Majumder}},\ and\ \bibinfo {author}
  {\bibfnamefont {P.}~\bibnamefont {Schmelcher}},\ }\href
  {https://doi.org/10.1103/PhysRevA.102.033320} {\bibfield  {journal} {\bibinfo
   {journal} {Physical Review A}\ }\textbf {\bibinfo {volume} {102}},\ \bibinfo
  {pages} {033320} (\bibinfo {year} {2020})}\BibitemShut {NoStop}%
\bibitem [{\citenamefont {Kwon}\ \emph {et~al.}(2021)\citenamefont {Kwon},
  \citenamefont {Mukherjee}, \citenamefont {Huh}, \citenamefont {Kim},
  \citenamefont {Mistakidis}, \citenamefont {Maity}, \citenamefont
  {Kevrekidis}, \citenamefont {Majumder}, \citenamefont {Schmelcher},\ and\
  \citenamefont {Choi}}]{kwon_2021_spontaneous}%
  \BibitemOpen
  \bibfield  {author} {\bibinfo {author} {\bibfnamefont {K.}~\bibnamefont
  {Kwon}}, \bibinfo {author} {\bibfnamefont {K.}~\bibnamefont {Mukherjee}},
  \bibinfo {author} {\bibfnamefont {S.~J.}\ \bibnamefont {Huh}}, \bibinfo
  {author} {\bibfnamefont {K.}~\bibnamefont {Kim}}, \bibinfo {author}
  {\bibfnamefont {S.~I.}\ \bibnamefont {Mistakidis}}, \bibinfo {author}
  {\bibfnamefont {D.~K.}\ \bibnamefont {Maity}}, \bibinfo {author}
  {\bibfnamefont {P.~G.}\ \bibnamefont {Kevrekidis}}, \bibinfo {author}
  {\bibfnamefont {S.}~\bibnamefont {Majumder}}, \bibinfo {author}
  {\bibfnamefont {P.}~\bibnamefont {Schmelcher}},\ and\ \bibinfo {author}
  {\bibfnamefont {J.-y.}\ \bibnamefont {Choi}},\ }\href
  {https://doi.org/10.1103/PhysRevLett.127.113001} {\bibfield  {journal}
  {\bibinfo  {journal} {Physical Review Letters}\ }\textbf {\bibinfo {volume}
  {127}},\ \bibinfo {pages} {113001} (\bibinfo {year} {2021})}\BibitemShut
  {NoStop}%
\bibitem [{\citenamefont {Ota}\ and\ \citenamefont
  {Astrakharchik}(2020)}]{ota_2020_leehuangyang}%
  \BibitemOpen
  \bibfield  {author} {\bibinfo {author} {\bibfnamefont {M.}~\bibnamefont
  {Ota}}\ and\ \bibinfo {author} {\bibfnamefont {G.}~\bibnamefont
  {Astrakharchik}},\ }\href {https://doi.org/10.21468/SciPostPhys.9.2.020}
  {\bibfield  {journal} {\bibinfo  {journal} {SciPost Physics}\ }\textbf
  {\bibinfo {volume} {9}},\ \bibinfo {pages} {020} (\bibinfo {year}
  {2020})}\BibitemShut {NoStop}%
\bibitem [{\citenamefont {Hu}\ and\ \citenamefont
  {Liu}(2020)}]{hu_2020_consistent}%
  \BibitemOpen
  \bibfield  {author} {\bibinfo {author} {\bibfnamefont {H.}~\bibnamefont
  {Hu}}\ and\ \bibinfo {author} {\bibfnamefont {X.-J.}\ \bibnamefont {Liu}},\
  }\href {https://doi.org/10.1103/PhysRevLett.125.195302} {\bibfield  {journal}
  {\bibinfo  {journal} {Physical Review Letters}\ }\textbf {\bibinfo {volume}
  {125}},\ \bibinfo {pages} {195302} (\bibinfo {year} {2020})}\BibitemShut
  {NoStop}%
\bibitem [{\citenamefont {Crank}\ and\ \citenamefont
  {Nicolson}(1947)}]{crank_nicolson_1947}%
  \BibitemOpen
  \bibfield  {author} {\bibinfo {author} {\bibfnamefont {J.}~\bibnamefont
  {Crank}}\ and\ \bibinfo {author} {\bibfnamefont {P.}~\bibnamefont
  {Nicolson}},\ }\href {https://doi.org/10.1017/S0305004100023197} {\bibfield
  {journal} {\bibinfo  {journal} {Mathematical Proceedings of the Cambridge
  Philosophical Society}\ }\textbf {\bibinfo {volume} {43}},\ \bibinfo {pages}
  {50–67} (\bibinfo {year} {1947})}\BibitemShut {NoStop}%
\bibitem [{\citenamefont {Lima}\ and\ \citenamefont
  {Pelster}(2011)}]{lima_2011_quantum}%
  \BibitemOpen
  \bibfield  {author} {\bibinfo {author} {\bibfnamefont {A.~R.~P.}\
  \bibnamefont {Lima}}\ and\ \bibinfo {author} {\bibfnamefont {A.}~\bibnamefont
  {Pelster}},\ }\href {https://doi.org/10.1103/PhysRevA.84.041604} {\bibfield
  {journal} {\bibinfo  {journal} {Phys. Rev. A}\ }\textbf {\bibinfo {volume}
  {84}},\ \bibinfo {pages} {041604} (\bibinfo {year} {2011})}\BibitemShut
  {NoStop}%
\end{thebibliography}%

\end{document}